\documentclass[twocolumn]{aa}  

\usepackage{color}
\usepackage{graphicx}
\usepackage[usenames,dvipsnames]{xcolor}
\usepackage{natbib}
\usepackage{txfonts}
\usepackage[hyperfootnotes=true, hidelinks, colorlinks, citecolor=blue, linkcolor=blue, linktocpage, bookmarks=true]{hyperref}
\usepackage{footnote}
\usepackage{float}

\usepackage[bottom, flushmargin]{footmisc}

\newcommand{\figps}[1]{\resizebox{\hsize}{!}{\rotatebox{0}{\includegraphics{#1}}}}

\makeatletter
\renewcommand*\aa@pageof{, page \thepage{} of \pageref*{LastPage}}
\makeatother

\begin{document}

\title{Modelling of the B-type binaries CW Cep and U Oph}
\subtitle{A critical view on dynamical masses, core boundary mixing, and core mass}

\author{C. Johnston\inst{1}
  \and K. Pavlovski\inst{2}
   \and A. Tkachenko\inst{1}}

\institute{Instituut voor Sterrenkunde, KU Leuven, Celestijnenlaan 200D, 3001, Leuven, Belgium, \email{colecampbell.johnston@kuleuven.be}
  \and Department of Physics, Faculty of Science, University of Zagreb, Bijeni\v{c}ka cesta 32, 10000 Zagreb, Croatia}

\date{Received Date Month Year / Accepted Date Month Year}

\abstract {Intermediate-Mass stars are often overlooked as they are not supernova progenitors but still host convective cores and complex atmospheres which require computationally expensive treatment. Due to this, there is a general lack of such stars modelled by state of the art stellar structure and evolution codes.} {This paper aims to use high-quality spectroscopy to update the dynamically obtained stellar parameters and produce a new evolutionary assessment of the bright B0.5+B0.5 and B5V+B5V binary systems CW Cep and U Oph.} {We use new spectroscopy obtained with the Hermes spectrograph to revisit the photometric binary solution of the two systems. The updated mass ratio and effective temperatures are incorporated to obtain new dynamical masses for the primary and secondary. With these, we perform isochrone-cloud based evolutionary modelling to investigate the core properties of these stars.} {We report the first abundances for CW\,Cep and U\,Oph as well as report an updated dynamical solution for both systems. We find that we cannot uniquely constrain the  amount of core boundary mixing in any of the stars we consider. Instead, we report their core masses and compare our results to previous studies.} {We find that the per-cent level precision on fundamental stellar quantities are accompanied with core mass estimates to between $\sim5-15$\%. We find that differences in analysis techniques can lead to substantially different evolutionary modelling results, which calls for the compilation of a homogeneously analysed sample to draw inference on internal physical processes.}

\maketitle


\section{Introduction}
The model independent estimates of the absolute dimensions of and distances to stars provided by eclipsing binary systems serve as a fundamental calibrator in modern astrophysics. In the best cases, such systems offer dynamical mass and radius estimates better than one per-cent \citep{torres2010}. Such precise measurements combined with the powerful constrains of co-evolution and identical initial chemical composition have allowed the thorough investigation of the importance of rotation in stellar evolutionary theory \citep{brott2011a,brott2011b,ekstrom2012,deMink2013,schneider2014,ekstrom2018}, the calibration of pre through post main-sequence evolution \citep{torres2013,higl2017,beck2018b,kirkbyKent2018}, the critical investigation of magnetic fields in stars \citep{takata2012,grunhut2013,torres2014a,pablo2015,kochukhov2018,wade2019}, the calibration of distances \citep{guinan1998,ribas2000c,ribas2005,hensberge2000,bonanos2006,pietrzynski2013,gallenne2016,suchomska2019}, the investigation of abundances and rotational velocities \citep{pavlovski2005,pavlovski2009a,pavlovski2009b,pavlovski2018,simondiaz2017}, as well as the calibration of asteroseismic modelling \citep{decat2000,decat2004,aerts2004,schmid2015,schmid2016,beck2018a,beck2018b,johnston2019}. Additionally, the advent of such precise measurements has led to the uncovering of the reported systematic discrepancy between masses obtained via dynamics or empirical spectral relations and fitting theoretically calculated evolutionary tracks \citep{herrero1992,ribas2000b,tkachenko2014}. This discrepancy has served as the centrepiece of intense debate over the importance of convective core boundary mixing in stellar evolution theory  \citep{ribas2000b,torres2010,torres2014b,tkachenko2014,stancliffe2015,claret2018,constantino2018,johnston2019}. 

In general, both element and angular momentum transport processes throughout a star are poorly calibrated \citep{aerts2019}. It is a well known short-coming of most 1-D theoretical descriptions of convection that convective boundaries are not well described \citep{hirschi2014}. Proposed as a means to remedy this short-coming, the inclusion of convective core overshooting as a way to increase near-core mixing in evolutionary models is now highly debated, with several competing studies claiming that models with and without overshooting can reproduce observed binaries across different mass ranges and evolutionary stages \citep{andersen1990,schroder1997,pols1997,claret2007,stancliffe2015,claret2016,claret2017,higl2017,constantino2018}. Convective core overshooting is a phenomenon theoretically predicted in intermediate- to high-mass stars with a convective core where the inertia of a convectively accelerated mass element propels said mass element beyond the convective boundary described by the Schwarzchild stability criterion into the stably stratified radiative region \citep{zahn1977,roxburgh1978,zahn1991,maeder2009}. The mathematical form of the implementation into stellar evolutionary codes is not universally agreed upon, with different forms having been successfully used to describe both binary  \citep{ribas2000b,guinan2000,tkachenko2014,claret2016,claret2017} and asteroseismic observations \citep{briquet2007,moravveji2015,moravveji2016,vanReeth2016,johnston2019}. To date, two such descriptions have been implemented in 1-D stellar evolution codes: i) convective penetration where the temperature gradient in the overshoot region is the adiabatic one, $\nabla_T=\nabla_{\rm ad}$, and ii) diffusive overshooting where the temperature gradient is the radiative one, $\nabla_T=\nabla_{\rm rad}$. This difference results in a fully chemically and thermally mixed extended region in the case of penetration, effectively meaning the core is extended and thus more massive. In the case of diffusive overshooting, the extended region is only partially chemically mixed, and any increase in core mass is due to the transport of chemicals into the convective core via this chemical mixing. In either case, the convective core will thus have more hydrogen available to burn (or He in the He-core burning phase), thus extending the main-sequence (MS) lifetime of the star, having a pronounced effect on the morphology of evolutionary tracks. Alternatively, some studies have used near-core rotational mixing to enhance the core mass, effectively producing the same situation where more rotational mixing corresponds to a more massive core. We point out that in 1D diffusive codes, the implementations of overshooting and rotational mixing are {seemingly different but both are} able to function as a proxy for the total amount of near-core mixing, whatever the physical cause, and that both prescriptions contain un-calibrated parameters. We adopt the approach of using overshooting as a general proxy for the total amount of near-core mixing, whatever its physical cause (rotation, convection, magnetism, waves). The mass-discrepancy reported between either spectroscopic \citep{herrero1992} or dynamical masses \citep{guinan2000,ribas2000b,claret2007,tkachenko2014} and evolutionary masses has traditionally been resolved by increasing the amount of overshooting in a stellar model. This increase in overshooting effectively increases the core mass at a given age, mimicking a more massive star. 

It was theoretically outlined that the extent of an overshooting region would be limited by the total energy (mass) of the core \citep{roxburgh1992}, and hence the mass of the star. This theoretical prediction has been investigated by numerous studies, some claiming no significant mass dependence \citep{schroder1997,pols1997,stancliffe2015,constantino2018}, while others claim a statistically significant mass dependence \citep{claret2007,claret2016,claret2017,claret2018,claret2019}. Yet another body of work suggests caution at the ability to constrain overshooting from classical observable quantities given the sensitivity of the data and methodologies \citep{valle2016,higl2017,valle2017,valle2018,johnston2019,constantino2018}. On the theoretical side, \cite{valle2016} studied the ability for models to uniquely describe a set of observables, revealing an inability to uniquely constrain overshooting. This result was supported by the findings of \cite{valle2018} and \cite{constantino2018} who show that traditional observed quantities do not provide enough discriminating power to uniquely constrain overshooting, with \cite{constantino2018} being unable to reproduce the mass-dependence of overshooting reported by \cite{claret2016}. Furthermore, \cite{johnston2019} showed that even with the inclusion of asteroseismic information, the extent of overshooting, stellar mass, and age cannot be uniquely constrained when properly accounting for correlated nature of stellar model parameters. Instead, \cite{johnston2019} suggest that the mass and radius of the convective core should be reported and considered in place of the overshooting.

In this paper, we follow the paradigm of \cite{johnston2019} to investigate the ability of well detached double-lined eclipsing binaries (EBs) to probe the core mass. Additionally, we investigate the comparatively sparsely sampled mass range of 4-6 \,M$_{\odot}$, connecting the mass ranges intensively covered by \cite{claret2016,claret2017} and \cite{pols1997,higl2017}. We revisit the intermediate- to high-mass double-lined EBs CW Cep and U Oph with new spectroscopy and radial velocities to obtain updated mass and radii estimates. In Section \ref{section:overview}, we will provide an overview of both systems, including past modelling efforts. In Sections \ref{section:spectra_orbit} and \ref{section:spectra_atmosphere} we discuss the new spectroscopy, the newly determined orbital elements from spectral disentangling, and the determination of spectroscopic quantities from the disentangled spectra, respectively. Section \ref{section:photometric_modelling} details the modelling procedure and results for both systems with the mass ratio fixed as derived in the previous section. Section~\ref{section:evol_models} covers our evolutionary modelling procedure. Sections \ref{section:discussion} and \ref{section:conclusions} discuss the newly determined mass and radii estimates for each system, the modelling results, and places them in the context of the larger modelling efforts of the community. Following the results of \cite{constantino2018} and \cite{johnston2019}, we report and discuss the estimated core mass and overshooting from our modelling procedure. 

\section{Literature overview on CW Cep and U Oph}
\label{section:overview}
\subsection{CW Cephei}

The detached double-lined EB CW Cephei (HD~218066, $V=7.6$\,mag) is an intensively studied system. The component masses have reported values ranging from $M_{1} = 11.82 - 13.49\,{\rm M_{\odot}}$ and $M_{2} = 11.09 - 12.05\,{\rm M_{\odot}}$ \citep{popper1974,popper1980,clausen_gimenez_1991,han2002}, placing this system at the lower end of the high-mass sequence. This spread in masses results in an uncertainty of $\sim 13$\% compared to the median value (solution $b$ by \cite{han2002}). The quality of the photometric light-curve solution, in particular the determination of the masses and radii, has been restricted by uncertainty in the mass and light ratio, respectively. This problem has been extensively discussed by \cite{clausen_gimenez_1991},who found that the spread in ratio of radii is also accompanied by a significant spread in the sum of the radii. Subsequent analysis of their own new photometry by \cite{han2002} and \cite{erdem2004} did not settle issue as they used a different methodology from \citet{clausen_gimenez_1991}. Namely, \cite{han2002} did not prefer the photographically determined light ratio over that returned from the lightcurve modelling, and \cite{erdem2004} allowed for the possibility of a-synchronous rotation in the components, which alters the light ratios derived from photometric modelling. Comparing the radii derived by different previous analyses (a complete set of the references are given in \cite{han2002}), a spread of  $\sim8$\% is found.

Apsidal motion was detected in CW\,Cep by \cite{nha1975} with improvements to the apsidal period made by \cite{han2002}, \cite{erdem2004}, and \cite{wolf2006}. The last authors settled the apsidal period to $U = 46.2\pm0.4$ yr, with an eccentricity of $e=0.0246$. The relatively short apsidal period, coupled with the brightness of the system have made it an ideal target for dynamical and evolutionary studies. Currently, the nature of the mechanism that drives the apsidal motion is not well understood. New space-based, high-precision, high duty-cycle observations from the NASA TESS mission \citep{ricker2015} promise to provide hitherto unseen constraints on the apsidal motion observed in this system.

Due to a distinct lack of constraints on the metallicity of CW\,Cep, the unique determination of evolutionary models for CW\,Cep has proven difficult \citep{clausen_gimenez_1991}. To date, several age estimates for CW\,Cep exist, with \cite{clausen_gimenez_1991} reporting an age of $\tau = (10\pm1)\, {\rm Myr}$, placing both components in the first half of the main-sequence. In their fitting work, \cite{ribas2000a} derived a much younger system with $\tau = 4.6\pm0.5\, {\rm Myr}$. However, \cite{ribas2000a} varied the metallicity and helium content, which introduces a near perfect degeneracy with age and mass in evolutionary modelling. Thus, their solution with $\tau = 4.6\pm0.5\, {\rm Myr}$, $Z = 0.023\pm0.007$, and a high helium contents albeit with a large uncertainty, $Y = 0.298\pm0.101$, is entirely consistent with that of \cite{clausen_gimenez_1991} given this degeneracy. Recently, CW Cep has been modelled by \cite{schneider2014}, who used a Bayesian modelling framework wrapped around Bonn evolutionary tracks to derive an age of $\sim 6 \, {\rm Myr}$, with best fit for an initial rotational velocity $v\sin i = 520$\,km\,s$^{-1}$ for both components. It should be noted that \cite{schneider2014} used a less massive solution in their modelling than \cite{ribas2000a} by $\sim0.5\, {\rm M_{\odot}}$ for the primary and $\sim0.3\, {\rm M_{\odot}}$ for the secondary and fixed the metallicity of their tracks to solar. Furthermore, \cite{blaauw1959} identified CW\,Cep to be a member of the Cep\,OB3, one of the smaller associations in the Orion arm. \cite{blaauw1961} also indicated that this association is composed of two subgroups. CW\,Cep is located in the older subgroup for which \cite{clausen_gimenez_1991} found an average age of about 10 Myr in perfect agreement with the age they obtained for CW\,Cep. However, in a comprehensive study of a new homogeneous $UBVRI$ photometry and membership \cite{jordi1996} obtained ages of 5.5 and 7.5 Myr for the two subgroups, in disagreement with the ages derived by both \cite{clausen_gimenez_1991} and \cite{schneider2014}.

CW\,Cep is also characterised as an intrinsically variable polarized object \citep{elias2008}. Both CW\,Cep and another early B+B binary system AH\,Cep, were observed with the Chandra X-ray Telescope in search for evidence of a wind-wind collision \citep{ignace2017}. Although CW\,Cep and AH\,Cep are comprised of stars with similar properties (c.f. \cite{pavlovski2018}), X-rays were only detected for AH\,Cep, despite it being nearly a factor 2 further away than CW\,Cep. The authors could not disentangle, however, whether the X-rays detected from AH\,Cep were caused by colliding winds, or perhaps from magnetic activity originating in one of the other components of the quadruple system of AH\,Cep \citep{ignace2017}.

\subsection{U Ophiuchi}

U\,Oph (HD~156247,$V=5.92$\,mag) is a detached double-lined EB comprised of two B5V components. Much like in the case of CW\,Cep, the dynamical solution of U\,Oph suffers from uncertainties in the light and mass ratios from spectral analysis. The reported masses for U\,Oph span from $M_{1} = 4.93 - 5.27\, {\rm M_{\odot}}$ and $M_{2} = 4.56 - 4.78\, {\rm M_{\odot}}$, whereas the reported radii span from $R_{1} = 3.29 - 3.48\, {\rm R_{\odot}}$ and $R_{2} = 3.01 - 3.11\, {\rm R_{\odot}}$ \citep{holmgren1991,vaz2007,wolf2006,budding2009}. This represents an uncertainty of $\sim7\%$ and $\sim5\%$ in $M_1$ and $M_2$ and an uncertainty of $\sim6\%$ and $\sim3\%$ in $R_1$ and $R_2$ when compared to the most recent solution by \cite{budding2009}. Additionally, a wide range of effective temperatures has been reported for both components, with differences up to 3\,000 K \citep{clements1979,eaton1973,holmgren1991,andersen1990,budding2009}.

A majority of the past lightcurve solutions are based on either the unfiltered photoelectric measurements of \cite{huffer1951}, the OAO-2 spacecraft photometery of \cite{eaton1973}, or both. However, the work of \cite{vaz2007} and \cite{budding2009} relies on new photometric and spectroscopic data. While all of these analyses used different modelling methodologies, codes, and assumptions, they produce derived quantities within a rather small range, as discussed above, and with high precision, which is promising.
U Oph displays a very rapid apsidal motion with a period of  $U\approx20$\,yr attributed to a distant third body \citep{koch1977,kaemper1986,wolf2002}. The apsidal motion has been studied thorougly with several proposed apsidal periods, some as large as 55 yr \citep{frieboes1973,panchatsaram1981,wolf2002,vaz2007}. Several recent studies have tried to constrain the nature of the tertiary component, reporting an orbital period of $P_{3}\approx38\,{\rm yr}$ and $M_3\approx1\,{\rm M_{\odot}}$ \citep{kaemper1986,wolf2002,vaz2007,budding2009}.

Largely due to uncertainties in its metallicity, there have been several discrepant ages reported for U\,Oph. \cite{holmgren1991} first reported an age of $\sim40$ Myr for U\,Oph when compared to evolutionary tracks without overshooting, and an age $\sim63$ Myr when compared to evolutionary tracks with overshooting. Later, \cite{vaz2007} compare their solution to evolutionary tracks of different metallicities, considering the apsidal constant as an additional constraint in their modelling and find the best agreement with isochrones for $\sim40$ Myr, $\sim52$ Myr, and $\sim62$ Myr calculated at $Z=0.02, 0.017$, and $0.01$, respectively. \cite{budding2009} perform their own evolutionary analysis, again with different codes and solutions compared to the previous evolutionary modelling attempts, and arrive at an average age estimate of $\sim38$ Myr between the two components for tracks calculated at $Z=0.02$. The authors also note that a younger solution is found at $\sim30$ Myr from tracks calculated at $Z=0.03$. Most recently, \cite{schneider2014} model U\,Oph with the Bonnsai code, assuming rotational mixing in their models (calculated at Z=0.014) and find an average age of $\sim41$ Myr for the system. \cite{budding2009} provide a comprehensive and detailed discussion of U Oph, to which we refer the reader for additional information.

\section{Orbital elements from new high-resolution spectroscopy}
\label{section:spectra_orbit}

For both CW\,Cep and U\,Oph, we obtained a new series of high-resolution \'{e}chelle spectra using the High Efficiency and high Resolution Mercator \'{E}chelle Spectrograph {\sc hermes} on the 1.2 m Mercator telescope at the Observatorio del Roque de los Muchachos, La Palma, Canary Islands, Spain. The {\sc hermes} spectrograph covers the entire optical and NIR wavelength range (3700 - 9100 {\AA}) with a spectral resolution of $R = 85\,000$ \citep{raskin2011}. CW\,Cep was observed a total of 18 times over 13 nights. Three observations were taken in January 2015 with the remaining 15 taken in August 2016. The argument of periastron progressed $\sim10^{\circ}$ between these two subsets, and less than one degree within either subset.  
U\,Oph, was observed 11 times over 10 nights from April to August 2016, during which time the argument of periastron progressed $\sim4^{\circ}$. The resulting spectra have an average S/N of 110 in a range 51-144 and 145 in a range 117-163 for CW\,Cep and U\,Oph, respectively.

The basic reduction of the spectra was performed with the {\sc hermes} pipeline software package. This pipeline delivers merged, un-normalised spectra. Therefore, before disentangling the spectra, we performed normalisation via spline function. 

\begin{table}
\caption{Orbital parameters determined by method of spectral disentangling. The periods were fixed from photometry in these calculations.}
\label{tab:SPD_orbit}
  \centering
  \begin{tabular}{lccc}
  \hline\hline
  Param.       &  Unit  &  CW\,Cep   &  U\,Oph  \\
  \hline
  $P$                 & d            & 2.72913159        &  1.67734590     \\
  $T_{\rm per}$       & d            & 57608.72$\pm$0.05 & -               \\
  $e$                 & -            & 0.0298$\pm$0.0008 & 0.              \\
  $\omega$            & deg          & 218.7$\pm$5.7     &  90.            \\
  $K_{\rm A}$         & km\,s$^{-1}$ & 211.1$\pm$0.4     & 181.1$\pm$0.6   \\
  $K_{\rm B}$         & km\,s$^{-1}$ & 230.2$\pm$0.4     & 200.6$\pm$0.8   \\
  \hline
  $q$                 &  -           & 0.917$\pm$0.002   & 0.903$\pm$0.005 \\
  $M_{\rm A}\sin^3 i$ &  M$_{\odot}$ & 12.66$\pm$0.05    & 5.08$\pm$0.04   \\
  $M_{\rm B}\sin^3 i$ &  M$_{\odot}$ & 11.61$\pm$0.04    &  4.59$\pm$0.04  \\
  $a\,\sin i$         &  R$_{\odot}$ & 23.78$\pm$0.03    &  12.65$\pm$0.03 \\
  
  \hline
  \end{tabular}
  \end{table}

Spectral disentangling (hereafter {\sc spd}) models the Doppler shift of spectral lines from a time-series of double-lined stellar spectra to determine the spectroscopic orbital elements as well as simultaneously reconstruct the individual spectra of the components \citep{simon1994}. Since the orbital elements are directly optimized in {\sc spd} the determination of radial velocities for each individual exposure is side-stepped. This removes the dependence on template spectra as are commonly used in the cross-correlation function (CCF) radial velocity (RV) determination method, which is often a source of systematic error due to mismatches between the spectral type of the star and that of the template \citep{hensberge2007}. Moreover, the resulting disentangled spectra of each component have an increased signal-to-noise compared to single-shot spectra, since disentangling acts as co-addition of the input spectra \citep[c.f.][]{pavlovski2010}. To perform {\sc spd}, we employ the {\sc FDBinary} code \citep{ilijic2004}, which performs {\sc spd} in Fourier space in order to efficiently solve the large and over-determined system of linear equations represented by the data, through applying discrete Fourier transforms to the spectra \citep{hadrava1995}. 

{\sc FDBinary} calculates the RV curve for each component using the standard set of orbital elements: period $P_{orb}$, time of periastron passage $T_{\rm per}$, eccentricity $e$, the argument of periastron $\omega_0$, and the semi-amplitudes of the RVs variations for the components $K_1$, and $K_2$. {\sc FDBinary} simultaneously optimises all orbital parameters across the entire set of spectra utilising the {\sc simplex} algorithm. Although the Balmer lines dominate the optical spectra of hot stars, these lines are broad and usually cover a majority of a single {\'e}chelle order, thus, any imperfections in the order-merging and normalisation procedure would propagate into the optimisation and affect both the orbital elements and the resulting disentangled component spectra. Therefore helium and metal lines are more suitable for our purposes. The resulting optimised orbital parameters for CW\,Cep and U\,Oph are listed in Table~\ref{tab:SPD_orbit}.

The orbital parameters of CW\,Cep and U\,Oph have been derived from fitting RVs in numerous previous studies. For CW\,Cep, \cite{strickland1992} determined $K_{\rm A} = 210.6\pm1.3$ km\,s$^{-1}$, and $K_{\rm B} = 229.9\pm1.6$ km\,s$^{-1}$ ($q = 0.92\pm0.1$) by fitting RVs extracted from three days of IUE spectra using a CCF method. However, since they had a relatively small number of spectra (21), the authors chose to fix the eccentricity to $e=0.0293$ following \cite{clausen_gimenez_1991}. Alternatively, \cite{popper1991} obtained $K_{\rm A} = 210\pm2$\,km\,s$^{-1}$  and $K_{\rm B} = 235\pm2$\,km\,s$^{-1}$ ($q=0.89\pm0.01$) by fitting RVs obtained via the CCF method from digitised plate spectra of CW\,Cep obtained with the Lick Observatory 3m telescope. The value for $K_{\rm B}$ derived by \cite{popper1991} is substantially larger than that obtained by \cite{popper1974} who used the very same data, but employed the oscilloscopic method to determine RVs as opposed to the CCF method that was used by \cite{popper1991}. By comparison, our results for CW\,Cep, listed in Table~\ref{tab:SPD_orbit}, place our estimates within $1\sigma$ of the solution presented by \cite{strickland1992} and within $2\sigma$ of \cite{popper1991}.

\citet{popper1991} also re-fit the orbital parameters of U\,Oph on RVs determined via CCF from Lick Observatory plate spectra, reporting  $K_{\rm A} = 183\pm2.5$\, km\,s$^{-1}$, and $K_{\rm B} = 195\pm3$\,km\,s$^{-1}$ ($q = 0.94\pm0.02$). Additionally, \citet{holmgren1991} reported $K_{\rm A} = 182\pm1$\, km\,s$^{-1}$ and $K_{\rm B} = 197\pm1$\,km\,s$^{-1}$ ($q = 0.924\pm0.007$) from 31 RV measurements extracted via CCF from 31 Reticon spectra obtained at the DAO 1.2m telescope. Later, \cite{vaz2007} obtained slightly different estimates of $K_{\rm A} = 182.7\pm1.2$\, km\,s$^{-1}$ and $K_{\rm B} = 203.3\pm1.6$\, km\,s$^{-1}$ ($q = 0.90\pm0.01$) from 34 plate spectra obtained by the ESO 1.5m telescope. Until this work, the only results based on {\'e}chelle spectra were presented by \cite{budding2009} who found $K_{\rm A} = 180.0\pm1.3$\,km\,s$^{-1}$ and $K_{\rm B} = 202.7\pm1.2$\,km\,s$^{-1}$ ($q = 0.89\pm 0.01$) from 30 RV measurements determined via CCF from spectra obtained with the {\sc hercules} spectrograph attached to the 1m Canterbury University McLellan Telescope located at Mt. John University Observatory in New Zealand. Our results are in rough agreement with the literature values, but highlight the increased precision provided by {\sc spd} which inherently minimises uncertainties presented by line-blending and template-mismatches that plague CCF techniques. 

\begin{figure}
    \centering
    \includegraphics[width=4.4cm]{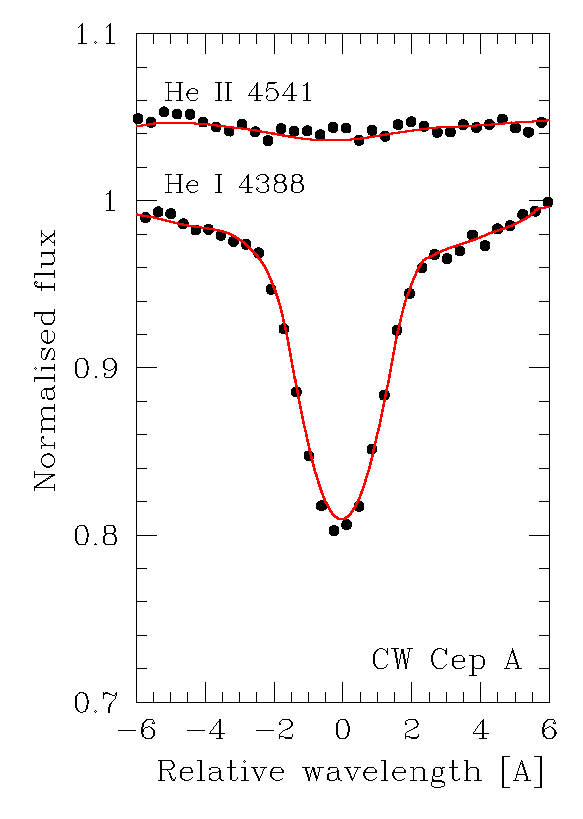}
    \includegraphics[width=4.4cm]{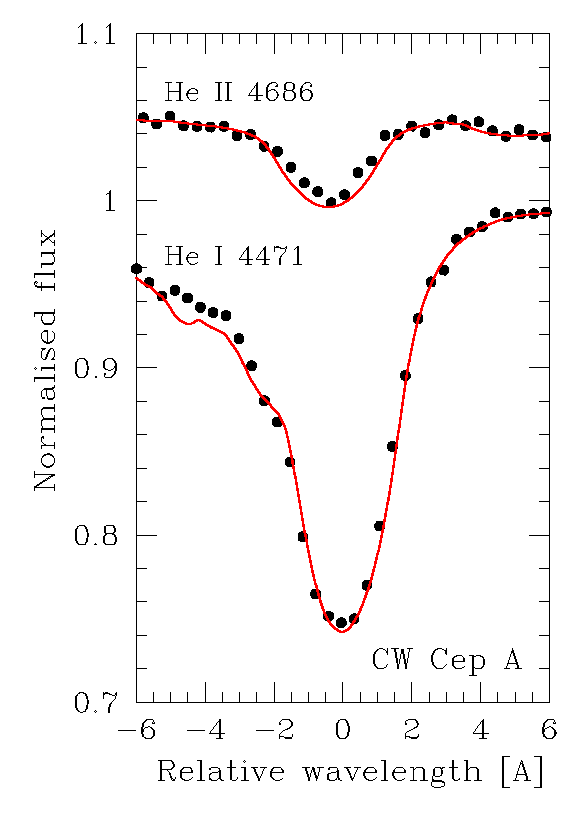}
    \includegraphics[width=4.4cm]{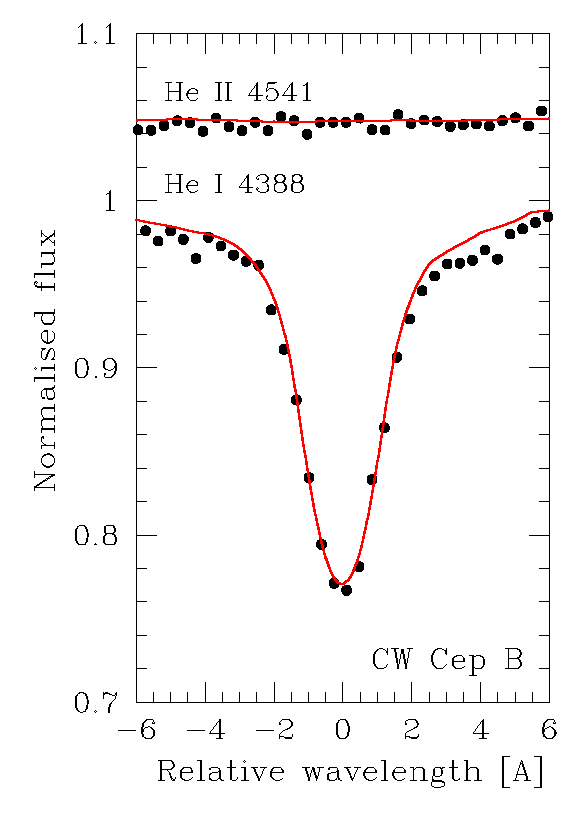}
    \includegraphics[width=4.4cm]{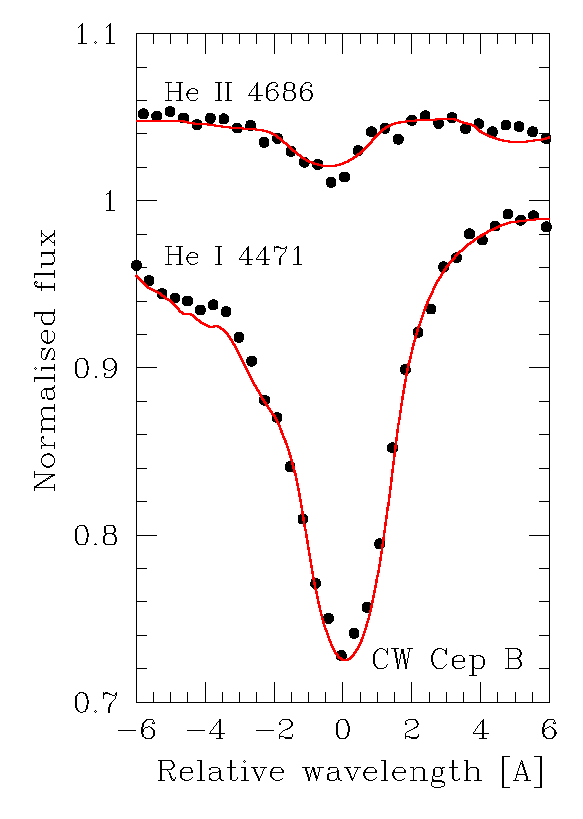}
    \caption{Determination of the $T_{\rm eff}$ for the components of CW\,Cep (the primary, component A, upper panels, the secondary. component B, bottom panels). The quality of fits are presented for \ion{He}{i} $\lambda$4388, and \ion{He}{ii} $\lambda$4541 lines (left column), and \ion{He}{i} $\lambda$4471, and \ion{He}{ii} $\lambda$4686 lines (right column). }
    \label{fig:cw_cep_spec_fit}
\end{figure}

\begin{table}
   \caption{Atmospheric parameters derived from an optimal fitting of re-normalised disentangled spectra for the components of CW\,Cep and U\,Oph. For CW\,Cep a grid of NLTE synthetic spectra was used, whilst
   for U\,Oph a grid of LTE synthetic spectra. The quantities given without the uncertainties were fixed in the calculation.}
    \centering
    \begin{tabular}{lcccc}
    \hline\hline
    Component    & $T_{\rm eff}$  &  $\log g$  &  $\xi_{\rm t}$  & $v\sin i$ \\
         &  [K]  & [dex]  & [km\,s$^{-1}$]  &  [km\,s$^{-1}$]   \\
    CW\,Cep A     &  28\,300$\pm$460 &  4.079 &  2.0$\pm$0.5 &  105.2$\pm$2.1    \\
    CW\,Cep B     &  27\,550$\pm$420 &  4.102 &  1.5$\pm$0.5 &  96.2$\pm$1.9    \\
    U\,Oph A      &  16\,580$\pm$180 &  4.073 &  2.0         &  110$\pm$6        \\
    U\,Oph B      &  15\,250$\pm$100 &  4.131 &  2.0         &  108$\pm$6        \\ 
    \hline
    \end{tabular}
    \label{tab:atmos_par}
\end{table}

\begin{table*}
\caption{Abundances determined for the components of binary system CW\,Cep. The atmospheric parameters used for the calculation of model atmospheres are given in Table~\ref{tab:atmos_par}. For the comparison the mean abundances for a sample of OB binaries given in Pavlovski et al.\ (2018), and for 'present-day cosmic standard' determined for a sample of a single  sharp-lined B-type stars in Nieva \& Przybilla (2012) are also presented.}
    \centering
    \begin{tabular}{lccccccc}
    \hline\hline
    Star  & C & N & O & [N/C] & [N/O] & Mg & Si  \\
    \hline
    CW Cep A  &  8.30 $\pm$ 0.07 & 7.79 $\pm$ 0.08 & 8.71 $\pm$ 0.07 & -0.51 $\pm$ 0.11 & -0.92 $\pm$ 0.11 & 7.55 $\pm$ 0.08 & 7.49 $\pm$ 0.06  \\
    CW Cep B  &  8.24 $\pm$ 0.07 & 7.70 $\pm$ 0.08 & 8.70 $\pm$ 0.06 & -0.54 $\pm$ 0.11 & -1.00 $\pm$ 0.10 & 7.53 $\pm$ 0.09 & 7.45 $\pm$ 0.07  \\
    \hline
    OB binaries &  8.26 $\pm$ 0.05 & 7.70 $\pm$ 0.04 & 8.71 $\pm$ 0.04 &  -0.56 $\pm$ 0.06 & -1.01 $\pm$ 0.06 & 7.59 $\pm$ 0.08 & 7.57 $\pm$ 0.10 \\
    B single stars & 8.33 $\pm$ 0.04 & 7.79 $\pm$ 0.04 & 8.76 $\pm$ 0.05 & -0.54 $\pm$ 0.06 &  -0.97 $\pm$ 0.06 & 7.56 $\pm$ 0.05 & 7.50 $\pm$ 0.05  \\
    \hline
    \end{tabular}
    
    \label{tab:cwcep_abund}
\end{table*}

\section{Atmospheric parameters from disentangled spectra}
\label{section:spectra_atmosphere}

\subsection{CW Cep}

CW\,Cep consists of two early-B spectral type stars with $T_{\rm eff} \sim28\,000$ K \citep{popper1974,popper1980,clausen_gimenez_1991,han2002}. These temperature estimates place both components in the temperature range where the strength of $\ion{He}{ii}$ lines starts to grow, thus allowing us to obtain precise effective temperature estimates through fine tuning the helium ionisation balance. As such, we apply the same methodology as described in \citet{pavlovski2018}, which we briefly summarise here. 

As an observed spectrum of a binary is a composite of spectra of the two components, the disentangled spectra are equal to the intrinsic components' spectra multiplied by the respective light factors, i.e. the components' fractional light contribution to the total light of a binary system, such that their co-addition reaches unity in the continuum. Generally, the fractional light contribution of each component can be determined either in the light curve analysis, or extracted from disentangled spectra. In the case of partially eclipsing binary systems where the components have similar radii, the light ratios are degenerate with the radii ratio and inclination. Therefore, it is advantageous to use the light ratio derived from disentangled spectra in the lightcurve modelling. We follow an iterative approach, where we first vary both the light factors and surface gravities, and then impose the light factors derived from spectroscopy as priors in our lightcurve modelling. To obtain atmospheric parameters, an optimised fit to the disentangled spectra of each component, which are re-normalised by their light-factor, is performed over a grid of pre-calculated non-local thermodynamic equilibrium (NLTE) models using the {\sc starfit} code \citep{tamajo2011,kolbas2014}. These theoretical NLTE spectra were calculated using {\sc Atlas9} model atmospheres and the NLTE spectral synthesis suite {\sc detail/surface} \citep{giddings1981,butler1984}. The synthetic spectra grid used in the optimisation contains models with $T_{\rm eff}\,\in\, 15\,000 - 32\,000 K$, and $\log g\,\in 3.5-4.5$ dex, and solar metallicity [M/H] = 0. However, we are able to fix the $\log g$ for each component according to the values listed in Table~\ref{tab:derived_pars}, since high precision, independent estimates for the surface gravities were derived from the light curve modelling. Fixing the surface gravity effectively lifts the degeneracy between the effective temperature and surface gravity, and enables us to use the Balmer lines as constraining information in our fit found by the helium ionisation balance. By fixing the surface gravity and micro-turbulence per component, we reduce the optimisation to eight free parameters: the effective temperature $T_{\rm eff}$ per component, projected rotational velocity $v\sin i$ per component, a relative Doppler shift between disentangled spectra, and laboratory reference frame, as well as the light-factors of the disentangled components. The optimisation across this parameter space is performed via a genetic algorithm modelled after that of the PIKAIA subroutine by \cite{charbonneau1995}, with the errors calculated via Markov Chain Monte Carlo technique as described by \cite{ivezic2014} and implemented by \cite{kolbas2014}. The optimisation was carried out over the spectral segment from 4000-4700 {\AA}, and includes the Balmer lines H$\gamma$ and H$\delta$, in addition to helium lines from both ionisation stages. Other spectral lines were masked. Due to the strong interstellar absorption band which effects the red wing of the $H\beta$ line, we were unable to use this spectral segment which covers the $y$ filter. However, since the effective temperatures of CW~Cep A \& B are similar, the wavelength dependence of the light-ratio is very small. The final analysis with fixed surface gravities and variable light ratios returned $T_{\rm eff,p} = 28\,300\pm460$ K, and $T_{\rm eff,s} = 27\,550\pm420$ K with light-factors of $0.565\pm0.005$ and $0.425\pm0.005$, for the primary and secondary, respectively. We note that these light factors are the same as those determined from the initial iteration, within errors The full optimised parameters are listed in Table \ref{tab:atmos_par}. The best fit for the $\ion{He}{i}$ and $\ion{He}{ii}$ lines for both components is shown in Fig.~\ref{fig:cw_cep_spec_fit}.

The reported values for the effective temperature of the primary of CW\,Cep have a broad range of almost 3\,000 K, from $T_{\rm eff,p} = 28\,000\pm1\,000$ K in \citet{clausen_gimenez_1991}, to a lower extreme $T_{\rm eff,p} = 25\,400$ in \cite{terrell1991}, and with intermediate values $T_{\rm eff,p} = 26\,500$ K in \citet{han2002} (\citet{terrell1991} and \citet{han2002} fix $T_{\rm eff,p}$ and do not report formal uncertainties for these values). It should be noted, however, that \citet{terrell1991} adopt their value for the primary effective temperature from a spectral type classification of B0.5, and \citet{clausen_gimenez_1991} determine a mean value from different color calibrated photometric relations. In these studies, the effective temperature of the secondary $T_{\rm eff,s}$ was then  determined from the light curve solution. The reported spread in secondary effective temperature is only 1\,300 K, with the hottest solution being only 600 K \citep{clausen_gimenez_1991} cooler than the primary and the coolest solution being 1\,900 K cooler than the primary \citep{terrell1991}. If we compare the values of our spectroscopically determined effective temperatures for the components of CW\,Cep, and the difference of their optimal values, $T_{\rm eff,p} = 28\,300\pm$460 K and $\Delta T_{\rm eff} = 750\pm620$ K, to the various estimates in previous analyses, we find the closest agreement with the estimates of \cite{clausen_gimenez_1991}. Comparatively, we are able to reduce the uncertainties considerably due to our methodology combining the spectral disentangling, ionisation balancing, and fixing the surface gravity.

Following our atmospheric analysis, we determine a detailed photospheric composition for both stars as well. We calculate {\sc ATLAS9} model atmospheres for the atmospheric parameters derived above, from which theoretical spectra are calculated with the {\sc detail/surface} suite. Details on the model atoms used can be found in \cite{pavlovski2018}. The abundances are then varied and optimised against the disentangled spectra, from which we report abundances for carbon, nitrogen, oxygen, magnesium, and silicon, as listed in Table \ref{tab:cwcep_abund}. Additionally, we are able to derive the microturbulence velocity $\xi_t$ from the oxygen-lines and the condition of nul-dependence of the oxygen abundance on equivalent width. The derived $\xi_t$ values for CW\,Cep\,A and B are listed in Table~\ref{tab:atmos_par}. For comparison, the 'present-day cosmic standard' abundance pattern for sharp-lined early-B type stars of \cite{nieva2012} is provided in the bottom row of Table \ref{tab:cwcep_abund}, with which we find general agreement. We also note that the abundances of CW\,Cep are in close agreement with the abundance pattern and ratios derived for OB binaries by \cite{pavlovski2018} as listed in the third row of Table \ref{tab:cwcep_abund}.

Since iron lines are not visible in early-B type stars, the iron abundance can not be directly measured and used as a proxy for stellar metallicity. Instead,  \cite{lyubimkov2005} determined the magnesium abundance from the \ion{Mg}{ii} line in a sample of 52 un-evolved early- to mid-B type stars and used this as a proxy for stellar metallicity. \cite{lyubimkov2005} determined the mean abundance $\log \epsilon({\rm Mg}) = 7.59\pm0.15$ to be in close agreement with the solar magnesium abundance, $\log \epsilon_{\odot}({\rm Mg}) = 7.55\pm0.02$ as determined in \cite{asplund+2009}. Exploiting the Mg abundance as a proxy for metallicity, \cite{lyubimkov2005} find that the metallicty of young MS B-type stars in the solar neighbourhood and the Sun are the same. Following this work, we find that our reported magnesium abundance suggests that CW\,Cep has solar metallicity. 

\begin{figure}
    \centering
    \includegraphics[width=8cm]{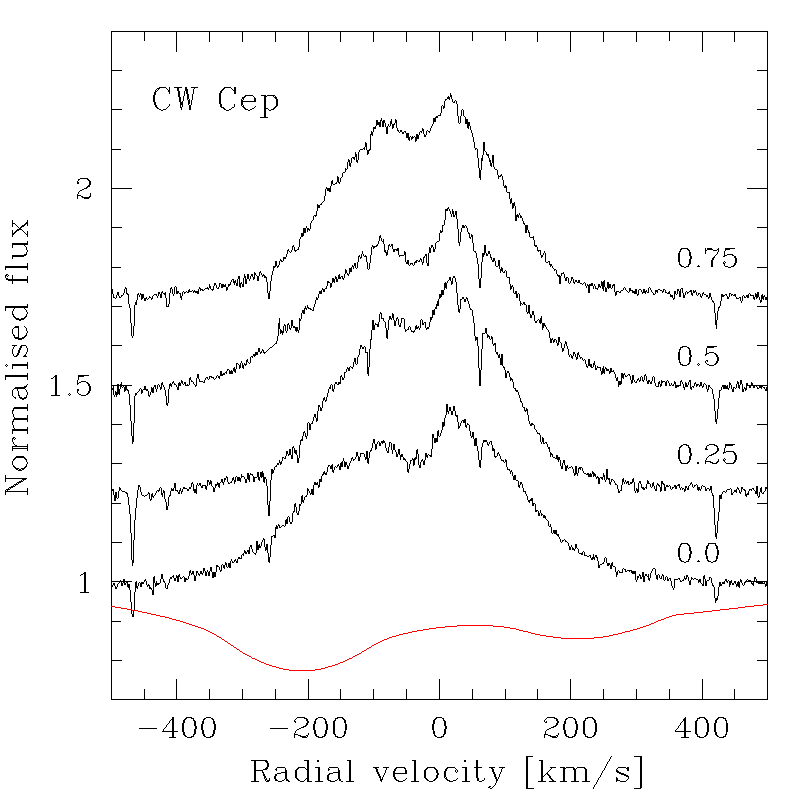}
    \caption{ Selected spectra corresponding to quarter phases centered around $H\alpha$ showing constant emission throughout the orbit. A synthetic composite spectrum of CW\,Cep at quarter phase is shown in red for comparison. Radial velocity is calculated in the rest-frame of the system.}
    \label{fig:cw_cep_halpha}
\end{figure}

Additionally, we note that we observe H$\alpha$ to be in emission in the new spectra assembled for this work. Fig.\,\ref{fig:cw_cep_halpha} displays spectra at roughly quarter phases as labelled, all of which show clear double-peaked emission with central absorption. The corresponding velocity difference between the blue ($V$) and red ($R$) peaks remains constant at $\sim105$\,km\,s$^{-1}$ through the orbital phase. Similarly, we find that the intensity ration between the two peaks remains roughly stable at $V/R \sim 0.95$ throughout the orbit as well. For comparison,  in Fig.\ \ref{fig:cw_cep_halpha} we show also synthetic H$\alpha$ profile for 0.25 phase. Although H$\alpha$ emission is typical for Be stars or mass-transfer binaries we cannot reliably attribute the emission to a given component. Moreover, as there is no clear evidence of variability of the emission with the orbital phase, we postulate that the emission originates from a circumbinary envelope or the nebula of the Cep~OB3 association in which CW\,Cep is located. An extensive \ion{H}{i} nebula in which the Cep~OB3 association is embeded is well documented \citep[e.g.][]{simonson1976}.

\subsection{U Oph}

U\,Oph consists of two main-sequence components of spectral type (mid-)B. Given that our NLTE grid discussed in the previous section is limited to stars hotter than 15\,000\,K, and that the use of the LTE formalism is overall justified for un-evolved stars in this temperature range, we employ the Grid Search in Stellar Parameters \citep[{\sc GSSP}, ][]{tkachenko2015} code for the analysis of the disentangled spectra of the U\,Oph's stellar components. The {\sc GSSP} algorithm is based on a grid search in basic atmospheric parameters ($T_{\rm eff}$, $\log g$, $\xi$, $v\sin i$, and [M/H]) and, if necessary, individual atmospheric abundances, and utilizes a $\chi^2$ merit function and statistics to judge the  goodness-of-fit between the grid of synthetic spectra and the observed spectrum and to compute 1$\sigma$ confidence intervals. Synthetic spectra are computed by means of the {\sc SynthV} radiative transfer code \citep{tsymbal1996} based on the pre-computed grid of {\sc LLmodels} atmosphere models \citep{shulyak2004}. Both atmosphere models and synthetic spectra can be computed for arbitrary chemical compositions, where one, several, or all chemical elements' abundances can be set, also with an option of a vertical stratification in the stellar atmosphere. Similarly, the effect of the microturbulent velocity can be taken into account, if necessary assuming its vertical stratification.

GSSP is a multi-function software for spectrum analysis that is able to deal with spectra of single stars ({\sc GSSP\_single} module), and those of spectroscopic double-lined binaries, either with their observed composite spectra ({\sc GSSP\_composite} module) or with the disentangled spectra of individual stellar components ({\sc GSSP\_single} or {\sc GSSP\_binary} module). In the former of the two binary cases ({\sc GSSP\_composite} module), a composite spectrum of a binary is fitted with a grid of composite synthetic spectra that are built from all possible combinations of grid points for the primary and secondary star. Individual radial velocities can also be optimised along with all the aforementioned atmospheric parameters of the two stars, where individual flux contributions are taken into account by means of the stellar radii ratio factor. In the latter case, the distinction is made whether the spectra are analysed as those of a single star with a certain light dilution factor ({\sc GSSP\_single} module, so-called unconstrained fitting where the light dilution factor is assumed to be independent of wavelength) or they are fitted simultaneously by optimising radii ratio to account for individual light contributions ({\sc GSSP\_binary} module, so-called constrained fitting with wavelength dependence of individual light contributions taken into account). A simultaneous fit of the two disentangled spectra is essential when a binary consists of two stars which are significantly different from each other in terms of their atmospheric properties. In this instance, their relative light contributions will strongly depend on wavelength. In the instance where the two stars have similar atmospheric parameters, independent fitting of the disentangled spectra is justified,  while still enforcing that the two (wavelength-independent) light factors ultimately add-up to unity \citep[see][for detailed discussion]{tkachenko2015}. 

\begin{figure}
    \includegraphics[width=9cm]{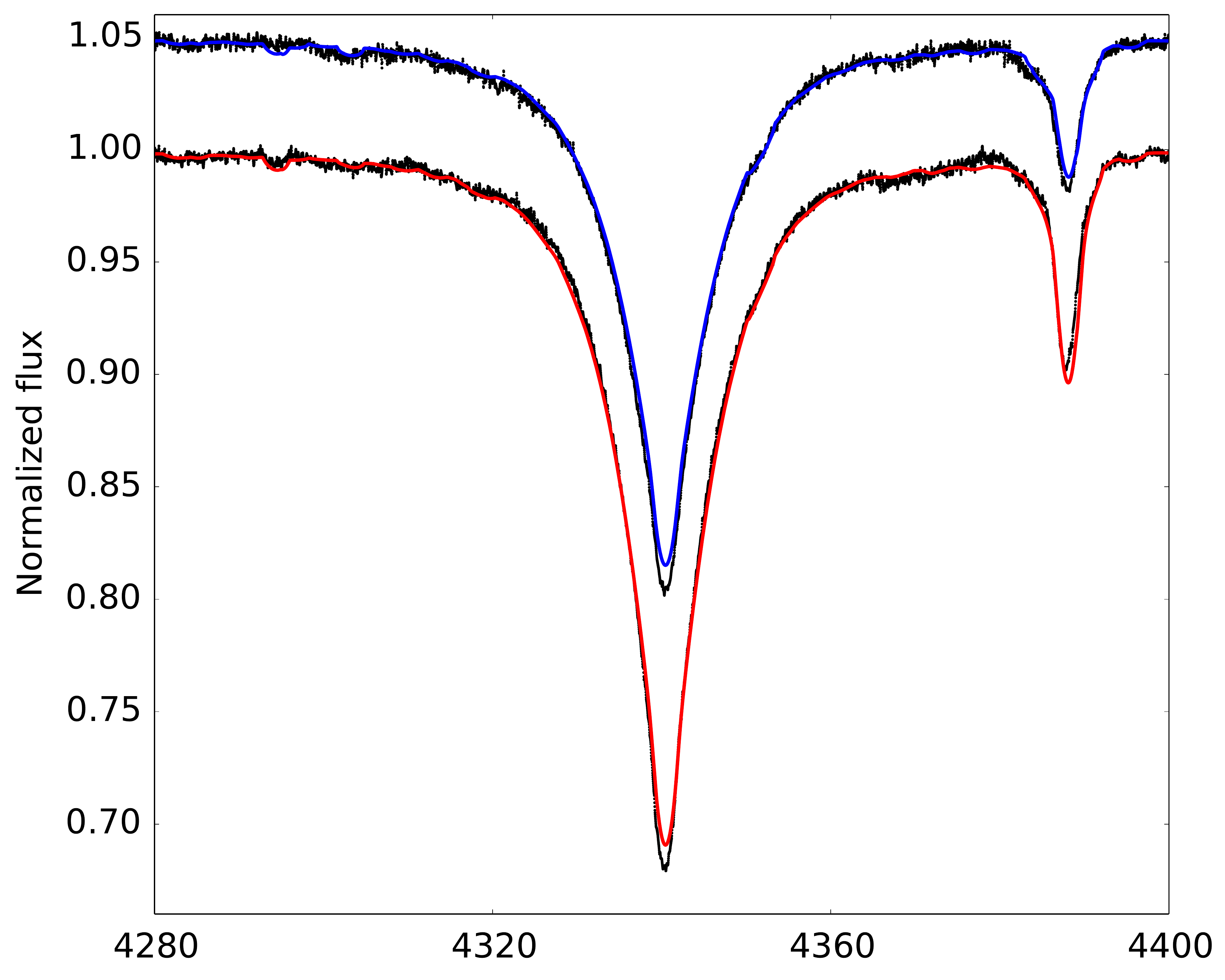}
    \includegraphics[width=9cm]{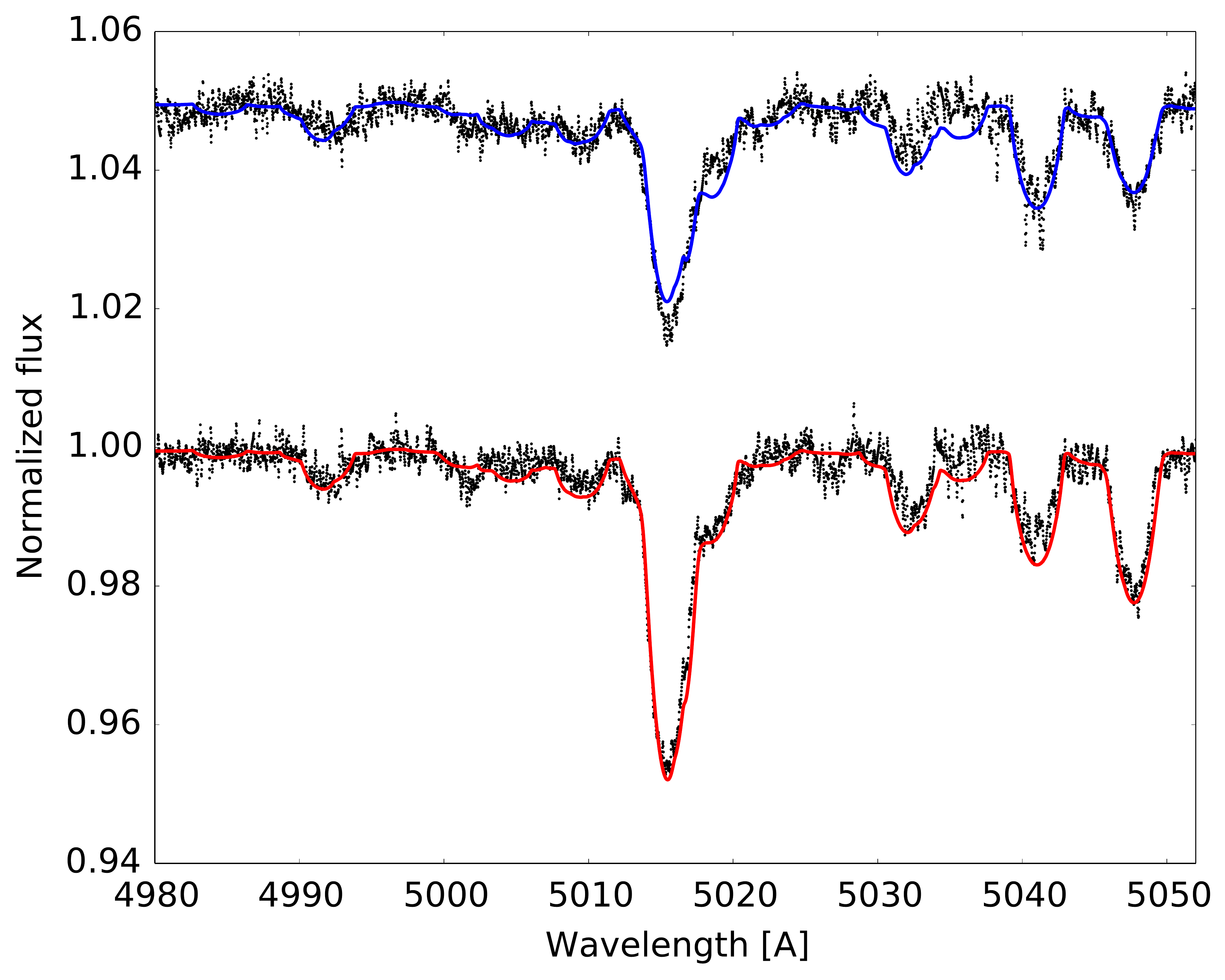}
    \caption{Quality of the fit of the disentangled (black dots) spectra with the synthetic spectra (red and blue solid line for the primary and secondary component, respectively) computed from the best fit parameters listed in Table~\ref{tab:atmos_par}. Spectra of the secondary component were vertically shifted by a constant factor for clarity.}
    \label{fig:u_oph_SpFit}
\end{figure}

As with CW~Cep, the atmospheric parameters of U~Oph A \& B are similar enough that we fit the disentangled spectra individually. Again, we use an iterative approach where the light-ratios are first determined from the disentangled spectra, then used as priors in the light curve solution. The photometric surface gravities are then fixed and the light-ratios are re-optimized along with the other atmospheric parameters from the disentangled spectra.  We found the light factors to be 0.575$\pm$0.007 and 0.425$\pm$0.008. The final solution is presented in Table~\ref{tab:atmos_par}, while the quality of the fit is demonstrated in Fig.~\ref{fig:u_oph_SpFit}.

\section{Revised Photometric Models}
\label{section:photometric_modelling}
Both CW Cep and U Oph have been studied extensively in the literature for several decades, with a heavy focus on the rapid apsidal motion displayed by the systems \citep{holmgren1991,clausen_gimenez_1991,han2002,wolf2002,erdem2004,wolf2006,vaz2007,budding2009}. This study aims to use the updated mass ratio, semi-major axis and effective temperatures of the primary and secondary obtained in Sections \ref{section:spectra_orbit} and \ref{section:spectra_atmosphere} to determine updated dynamical masses, radii, and surface gravities from photometric modelling with {\sc PHOEBE} \citep{prsa2005,prsa2011}.

\subsection{Photometric Data}

For CW Cep we revisit the photometry initially analysed by \cite{clausen_gimenez_1991}. These data consists of 21 nights of observations spanning 3 years in the Stromgren $uvby$ photometric system, totalling 1396 measurements in the $uby$ filters, and 1318 in the $v$ filter. Both HD 218342 and HD 217035 served as photometric comparison stars, from which the final differential magnitudes were obtained. Extinction corrections were applied to the data as were determined by nightly coefficients determined across the listed comparison stars and other standard objects \citep{gimenez1990}. According to \cite{clausen_gimenez_1991}, the observations were constant to 0.004 mag in all filters, which we adopt as the uncertainty on each point. 

We also revisit archival data for U Oph, initially analysed by \cite{vaz2007}. These data consist of 25 nights of observations spanning 1992-1994 in the Stromgren $uvby$ photometric system, totalling 645 measurements, however, due to a trend in the data, we do not use the $u$-band lightcurve. The data were taken with the 0.5m ESO SAT telescope in La Sille, Chile. HR 6367, HR 6353, and SAO 122251 were all used as comparison stars, from which the final differential magnitudes were obtained. As with CW Cep, extinction corrections were calculated each night from the comparison stars used. For more information on the comparison targets and observations, we refer to \cite{vaz2007}. Finally, \cite{vaz2007} report a standard deviation of 0.0037 mag in the $vby$ filters, which we adopt as the uncertainty on each point.

\subsection{Photometric modelling methodology}
Both CW Cep and U Oph are well detached systems, exhibiting mild out of eclipse variability and slow apsidal motion, which for the purposes of our modelling is effectively mitigated by phase-binning the data. Our photometric modelling uses the {\sc PHOEBE} binary modelling code, which is a modern extension of the original WD code but also incorporates new physics such as dynamic effects, the light travel time effect, and the reflection effect \citep{prsa2005,prsa2011}. Given that all components considered are expected to have radiative envelopes, we fix the gravity darkening exponent to unity for all components \citep{vonZeipel1924}. To obtain statistically robust estimates for the fit parameters, we wrap {\sc PHOEBE} into a  Bayesian Markov Chain Monte Carlo (MCMC) framework using the {\sc emcee} affine-invariant ensemble sampler MCMC code \citep{foremanMackey2013}, which has already been successfully applied by \cite{schmid2015,hambleton2016,pablo2017,johnston2017,kochukhov2018}.

MCMC procedures numerically evaluate Bayes' Theorem, given by:
\begin{equation}
p\left(\Theta|d\right)\propto\mathcal{L}\left(d|\Theta\right)\,p\left(\Theta\right), 
\end{equation}
\label{bayes}
to estimate the posterior probability, $p\left(\Theta|d\right)$, of some varied parameters $\Theta$ given the data $d$.We can see above that $p\left(\Theta|d\right)$ is proportional to the product of the likelihood function $\mathcal{L}\left(\Theta|d\right)$ and the prior probability of the parameter vector $p\left(\Theta\right)$. We write the likelihood function as:
\begin{equation}
\label{likelihood}
\mathrm{ln}\mathcal{L}\propto -\frac{1}{2}\sum_{i}\left( \frac{d_i - y\left(\Theta\right)_i}{\sigma_i} \right)^2 \, ,
\end{equation} 
where, $y\left( \Theta \right)_i$ is each individual model point and $\sigma_i$ are the individual uncertainties associated with the data. We have written the log-likelihood function above as this is what is used in practice. To make efficient use of the information obtained via the spectroscopic analysis,  we apply Gaussian priors on the light factors (per-cent contribution per component) and $v\sin i$ estimates per component, as well as the projected binary separation $a\sin i$, the mass ratio $q$, the effective temperature of the secondary $T_{\rm eff,2}$, and the eccentricity of the orbit. However, since PHOEBE does not directly sample all of these, we calculate the $v\sin i$ separately for each component and $a\sin i$ for every $\Theta$ considered. By including the spectroscopic light factors and simultaneously fitting all filters, we arrive at a more robust solution than if we were to fit them all individually and mitigate any degeneracies between the temperatures, light-factors, and potentials of each component \citep{clausen_gimenez_1991}. Furthermore, inclusion of priors on $v\sin i$ for each component helps constrain the spin paramters $f_1=\omega_{rot,1}/\omega_{orb}$ and $f_2=\omega_{rot,2}/\omega_{orb}$, which are otherwise largely unconstrained.

We draw parameter estimates and uncertainties as the median and $68.27\%$ ($1\sigma$) Highest Posterior Density (HPD) intervals of the marginalised posterior distribution for each sampled parameter. As both systems undergo apsidal motion, we bin each lightcurve such that each phase bin covers 0.0033 phase units, which covers the entire periastron advance in a single binned point for either system. Although PHOEBE accepts $e$ and $\omega$ directly, we sample $e\sin\omega$ and $e\cos\omega$ in our MCMC analysis and solve for $e$ and $\omega$ afterwards. To aid in the discussion and provide additional constraints, we also report the relative radii in the bottom panel of Table~\ref{tab:binary_fit_pars}.

\subsection{PHOEBE Model: CW Cep}

To propagate our newly derived spectroscopic and orbital information into updated dynamical masses and radii, we fix the effective temperature of the primary ($\mathrm{T_{eff_1}}$) to the value listed in Table~\ref{tab:atmos_par}. As mentioned above, we apply Gaussian priors on the mass ratio, the eccentricity, the projected binary separation, the $v\sin i$ per component, and the light-factor per component in the $v$- and $b$-band lightcurves, since these correspond to the spectral range for which we derived the light factors. The light factor for the $u$- and $y$-bands are given a uniform prior. For each sampled $\Theta$, we interpolate limb-darkening coefficients for the square-root law from the provided {\sc PHOEBE} girds. Finally, given the radiative envelope of hot stars such as CW Cep A \& B, we fix the albedo to unity in both components.

CW Cep is known to suffer from third light which scales the apparent eclipse depths across each filter. Accounting for this scaling is non-trivial as there is a degeneracy between inclination and third light levels. However, this degeneracy is crucial to account for when determining the derived masses. We use a uniform prior on the third light contributions per filter. Additionally, we sample the reference date ($\mathrm{HJD_0}$), period ($P_{\rm orb}$), the inclination, the total binary separation, the secondary effective temperature, as well as potentials and synchronicity parameters per component ($\Omega_{1,2}$ and $f_{1,2}$, respectively), giving all uniform priors. All sampled values, and their type of prior, are noted in Table \ref{tab:binary_fit_pars}.

The analysis of \cite{clausen_gimenez_1991} states that the argument of periastron changes $24^{\circ}$ from $\sim287^{\circ}$ to $\sim311^{\circ}$ over the course of the photometric campaign, which corresponds to the secondary minima shifting $\sim0.007$ phase units. To mitigate this change in periastron, we phase bin our data to 300 points, with each bin covering 0.0033 phase. Thus, the argument of periastron that we sample does not correspond to the value provided in the literature of the zero-point, but rather to the mean periastron during the photometric campaign.

\begin{figure}
\centering
\includegraphics[width=\linewidth]{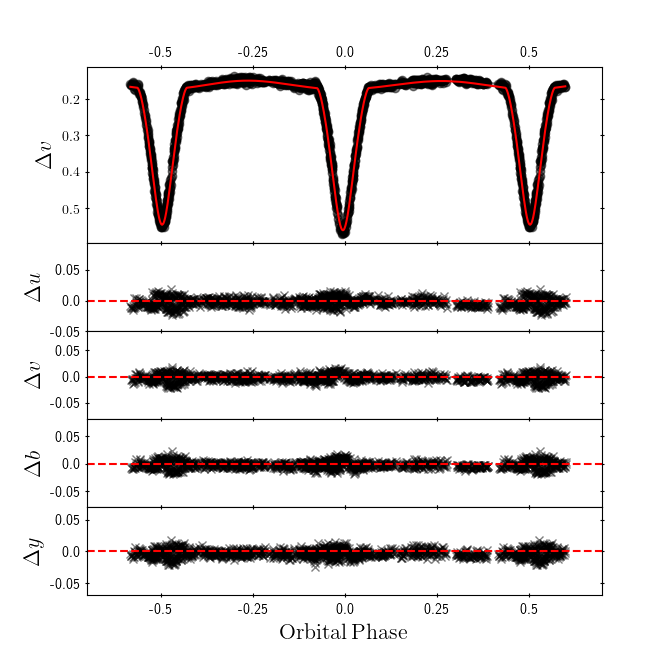}
\caption{ Top panel displays the CW\,Cep PHOEBE Model (solid red) for the Stromgren {\it v} lightcurve (black x's) constructed from median values reported in Table \ref{tab:binary_fit_pars}. Lower panels show the residual lightcurves in the $uvby$ filters after the best model has been removed. The dashed red-line denotes the zero-point to guide the eye.}
\label{fig:cw_cep_lc_stromgren}
\end{figure}

The third column of Table \ref{tab:binary_fit_pars} shows the median and HPD estimates for the best fitting model. These values were used to construct the models seen in Fig. \ref{fig:cw_cep_lc_stromgren}. 

For a consistency check, we compare the luminosities derived from the binary modelling with the luminosity derived from the Gaia parallax for CW~Cep: $\pi_G=1.04\pm0.49$ mas \citep{luri2018,lindegren2018}. We take $A_v=1.96$ following the reported value of ${\rm E}(b-y)$ from \citet{clausen_gimenez_1991} and $BC_v = 2.95\pm0.05$ calculated as the average correction between CW~Cep A and B \citep{reed1998}. The summed luminosity derived from our binary model yields $\log\frac{L}{L_{\odot}}=4.48\pm0.02$, while the GAIA derived luminosity yields $\log\frac{L_G}{L_{\odot}}=4.78\pm0.41$. Given the large ($\sim50\%$ uncertainty on the Gaia parallax, we also check the luminosity derived from the Hipparcos parallax \citep[$\pi_H=1.57\pm0.69$ mas;][]{vanLeeuwen2007} which yields  $\log\frac{L_H}{L_{\odot}}=4.42\pm0.4$. We find that all of these agree within the uncertainties.

\subsection{PHOEBE Model: U Oph}

As with CW\,Cep, we fix the effective temperature of the primary to the value listed in Table~\ref{tab:atmos_par} and impose Gaussian priors on $q$, $a\sin i$, $e$, $v\sin i$ per component and the light factors per component in the $v$- and $b$-band lightcurves. Although we can safely ignore the small eccentricity and set it to zero to perform {\sc spd}, we cannot ignore the eccentricity in the lightcurve. As such, we apply a Gaussian prior according to the values taken from \citet{vaz2007}. Limb-darkening coefficients are interpolated from {\sc PHOEBE} tables at every model evaluation. The albedos of both components are fixed to unity as both stars are expected to have radiative envelopes.

\begin{figure}
\centering
\includegraphics[width=\linewidth]{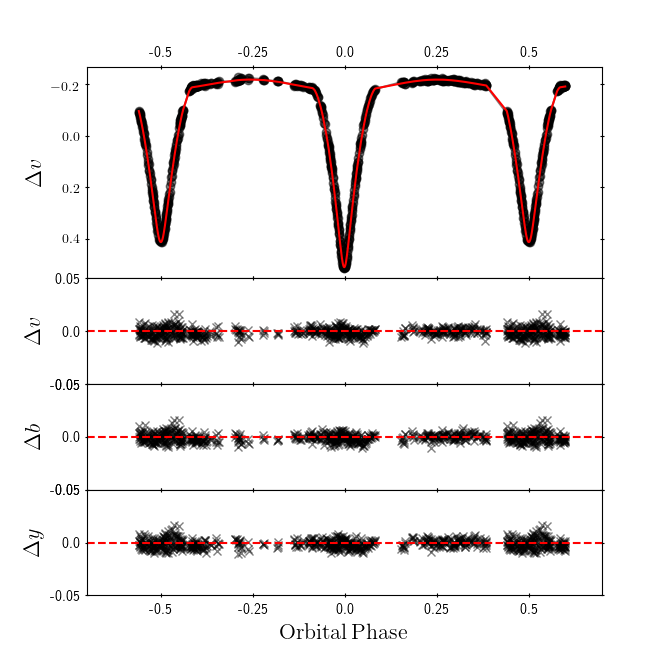}
\caption{Top panel displays the U~Oph PHOEBE Model (solid red) for the Stromgren {\it v} lightcurve (black x's) constructed from median values reported in Table \ref{tab:binary_fit_pars}. Lower panels show the residual lightcurves in the $vby$ filters after the best model has been removed. The dashed red-line denotes the zero-point to guide the eye.}
\label{fig:u_oph_lc}
\end{figure}

Since U~Oph is also known to suffer from third light, we take the same approach as with CW Cep, using  uniform priors for the third light per filter and uniform priors in all other parameters listed in Table \ref{tab:binary_fit_pars}. To mitigate the effects of the apsidal advance, we phase bin into 300 bins, which effectively covers the apparent change in superior / inferior conjunction. Again, this means that the argument of periastron reported is an average over the photometric campaign when the data was collected. The best model according to the median estimates listed in Table~\ref{tab:binary_fit_pars} is shown in Fig.~\ref{fig:u_oph_lc}. Derived parameters for both CW\,Cep and U\,Oph are reported in Table~\ref{tab:derived_pars} alongside other solutions from the literature. 

As with CW~Cep, we compare the total luminosity obtain from binary modelling with the luminosities derived from GAIA \citep[$\pi_G=3.74\pm0.13$ mas;][]{luri2018,lindegren2018} and Hipparcos \citep[$\pi_H=4.99\pm0.41$ mas][]{vanLeeuwen2007}, assuming $A_v=0.72\pm0.2$ as taken from \citet{vaz2007} . The luminosity we calculate as $\log\frac{L}{L_{\odot}}=3.12\pm0.01$ does not agree with the GAIA derived luminosity as $\log\frac{L_G}{L_{\odot}}=3.27\pm0.09$ within $1\sigma$, but does agree with the Hipparcos derived luminosity as $\log\frac{L}{L_{\odot}}=3.02\pm0.1$ within $1\sigma$.

\begin{table*}
\caption{\label{tab:binary_fit_pars} Binary model parameters for CW\,Cep and U\,Oph.}
\centering
\begin{tabular}{lllll}
\hline\hline
& \multicolumn{2}{c}{CW\,Cep} & \multicolumn{2}{c}{U\,Oph}\\
Parameter&Prior &HPD Estimate&Prior &HPD Estimate \\
\hline
\multicolumn{5}{c}{Sampled Parameters} \\
\hline
\rule{0pt}{2.3ex}${\rm L_{1,u}\,[\%]}$&$\mathcal{U}\left(40,70\right)$&$57.3\substack{+1.1 \\ -1.0}$& -- & -- \\
\rule{0pt}{2.2ex}${\rm L_{1,v}\,[\%]}$&$\mathcal{N}\left(56.5,0.5\right)$&$56.7\substack{+1.1 \\ -1.0}$&$\mathcal{N}\left(57.5,0.7\right)$&$57.2\substack{+2.3 \\ -5.2}$\\
\rule{0pt}{2.2ex}${\rm L_{1,b}\,[\%]}$&$\mathcal{N}\left(56.5,0.5\right)$&$56.6\substack{+1.1 \\ -1.0}$&$\mathcal{N}\left(57.5,0.7\right)$&$57.1\substack{+2.2 \\ -5.1}$\\
\rule{0pt}{2.2ex}${\rm L_{1,y}\,[\%]}$&$\mathcal{U}\left(40,70\right)$&$56.5\substack{+1.1 \\ -1.0}$&$\mathcal{U}\left(40,70\right)$&$57.0\substack{+2.2 \\ -5.2}$\\
\rule{0pt}{2.3ex}${\rm L_{3,u}\,[\%]}$&$\mathcal{U}\left(0,15\right)$&$0.6\substack{+0.6 \\ -0.5}$& -- & --\\
\rule{0pt}{2.2ex}${\rm L_{3,v}\,[\%]}$&$\mathcal{U}\left(0,15\right)$&$1.9\substack{+0.6 \\ -0.5}$&$\mathcal{U}\left(0,15\right)$&$0.8\substack{+0.2 \\ -0.3}$\\
\rule{0pt}{2.2ex}${\rm L_{3,b}\,[\%]}$&$\mathcal{U}\left(0,15\right)$&$2.8\substack{+0.5 \\ -0.5}$&$\mathcal{U}\left(0,15\right)$&$1.1\substack{+0.2 \\ -0.2}$\\
\rule{0pt}{2.2ex}${\rm L_{3,y}\,[\%]}$&$\mathcal{U}\left(0,15\right)$&$3.6\substack{+0.6 \\ -0.4}$&$\mathcal{U}\left(0,15\right)$&$1.3\substack{+0.2 \\ -0.2}$\\
\rule{0pt}{2.2ex}${\rm T_{eff,s}\,[K]}$&$\mathcal{N}\left(27550,600\right)$&$27420\substack{+150 \\ -120}$&$\mathcal{N}\left(15620,200\right)$&$15820\substack{+90 \\ -90}$\\
\rule{0pt}{2.2ex}$P_{\rm orb}\,{\rm [d]}$&$\mathcal{U}\left(1,5\right)$&$2.7291316\substack{+4e-7 \\ -3e-7}$&$\mathcal{U}\left(1,4\right)$&$1.67734590\substack{+2e-8 \\ -2e-8}$\\
\rule{0pt}{2.2ex}${\rm HJD_0\,[d]}$&$\mathcal{U}\left(-2,2\right)+2441669$  &$0.5831\substack{+0.0005 \\ -0.0006}$&$\mathcal{U}\left(-2,2\right)+2449161$&$0.61101\substack{+0.00003 \\ -0.00002}$\\
\rule{0pt}{2.2ex}$i\,{\rm [\deg]}$&$\mathcal{U}\left(70,90\right)$ &$81.804\substack{+0.006 \\ -0.004}$&$\mathcal{U}\left(70,90\right)$&$87.86\substack{+0.1 \\ -0.08}$\\ 
\rule{0pt}{2.3ex}$e\sin\omega_0$&$\mathcal{U}\left(-0.0287,0.0287\right)$ &$-0.02544\substack{+2e-5 \\ -2e-5}$& $\mathcal{U}\left(-0.003,0.003\right)$ & $0.00189\substack{+1e-5\\-1e-5}$\\ 
\rule{0pt}{2.3ex}$e\cos\omega_0$&$\mathcal{U}\left(-0.0287,0.0287\right)$ &$0.01329\substack{+4e-5 \\ -3e-5}$& $\mathcal{U}\left(-0.003,0.003\right)$ & $0.00233\substack{+1e-5\\-1e-5}$\\ 
\rule{0pt}{2.2ex}$a\,{\rm [R_{\odot}]}$&$\mathcal{U}\left(5,40\right)$&$24.01\substack{+0.04 \\ -0.04}$&$\mathcal{U}\left(5,40\right)$&$12.66\substack{+0.03 \\ -0.03}$\\ 
\rule{0pt}{2.2ex}$q=\frac{M_2}{M_1}$&$\mathcal{N}\left(0.92,0.002\right)$&$0.919\substack{+0.005 \\ -0.005}$&$\mathcal{N}\left(0.90,0.01\right)$&$0.90\substack{+0.01 \\ -0.01}$\\ 
\rule{0pt}{2.2ex}${\rm \Omega_{1}}$&$\mathcal{U}\left(4.5,9\right)$ &$5.39\substack{+0.05 \\ -0.03}$&$\mathcal{U}\left(4,9\right)$&$4.64\substack{+0.02 \\ -0.02}$\\
\rule{0pt}{2.2ex}${\rm \Omega_{2}}$&$\mathcal{U}\left(4.5,9\right)$ &$5.43\substack{+0.04 \\ -0.05}$&$\mathcal{U}\left(4,9\right)$&$4.84\substack{+0.05 \\ -0.05}$\\
\rule{0pt}{2.2ex}$f_1$&$\mathcal{U}\left(0.5,2\right)$ &$1.06\substack{+0.03 \\ -0.03}$& $\mathcal{U}\left(0.5,2\right)$ & $1.07\substack{+0.07 \\ -0.07}$ \\
\rule{0pt}{2.2ex}$f_2$&$\mathcal{U}\left(0.5,2\right)$ &$1.03\substack{+0.03 \\ -0.03}$& $\mathcal{U}\left(0.5,2\right)$ & $1.16\substack{+0.08 \\ -0.07}$ \\ 

\hline
\multicolumn{5}{c}{Geometric Parameters} \\
\hline
\rule{0pt}{2.2ex}$r_1$& &$0.227\substack{+0.001 \\ -0.002}$&  & $0.2715\substack{+0.0005 \\ -0.0005}$ \\
\rule{0pt}{2.2ex}$r_2$& &$0.212\substack{+0.002 \\ -0.001}$&  & $0.2408\substack{+0.0007 \\ -0.0009}$ \\

\hline
\end{tabular}
\tablefoot{
The top panel shows those parameters which were sampled during the MCMC run. For each parameter we list the units, when applicable, the priors, and the estimated values from the median and HPD confidence intervals. The bottom panel displays derived geometric parameters and their estimates. Gaussian priors are listed with an $\mathcal{N}$, followed by their mean and width, and uniform priors are listed with a $\mathcal{U}$,  followed by their boundaries.
}
\end{table*}

\begin{table*}
\caption{\label{tab:derived_pars} Derived Parameters CW Cep \& U Oph.}
\centering
\begin{tabular}{lccccc}
\hline\hline
CW Cep \\
\hline
Parameter &\cite{gimenez1987} &\cite{clausen_gimenez_1991} &\cite{han2002}\tablefootmark{a} &\cite{han2002}\tablefootmark{b} &This Work\\
\rule{0pt}{2.4ex}$\mathrm{M_1\,[M_{\odot}]}$  &$11.9\pm0.1$ &$11.82\pm0.14$ &$13.49$ &$12.93$ &$13.00\substack{+0.07 \\ -0.07}$\\
\rule{0pt}{2.2ex}$\mathrm{M_2\,[M_{\odot}]}$  &$11.2\pm0.1$ &$11.09\pm0.14$ &$12.05$ &$11.84$ &$11.94\substack{+0.08 \\ -0.07}$\\
\rule{0pt}{2.2ex}$\mathrm{R_1\,[R_{\odot}]}$  &$5.40\pm0.1$ &$5.48 \pm0.12$ &$6.03 $ &$5.97 $ &$5.45 \substack{+0.03 \\ -0.06}$\\
\rule{0pt}{2.2ex}$\mathrm{R_2\,[R_{\odot}]}$  &$4.95\pm0.1$ &$4.99 \pm0.12$ &$4.60 $ &$4.56 $ &$5.09 \substack{+0.06 \\ -0.03}$\\ 
\rule{0pt}{2.2ex}$\log g_1 \, \mathrm{[dex]}$ &$4.05\pm0.02$&$4.03 \pm0.02$ &$4.01 $ &$3.99 $ &$4.079 \substack{+0.010 \\ -0.005}$\\
\rule{0pt}{2.2ex}$\log g_2 \, \mathrm{[dex]}$ &$4.10\pm0.02$&$4.09 \pm0.02$ &$4.19 $ &$4.19 $ &$4.102 \substack{+0.005 \\ -0.010}$\\
\hline

\hline
U Oph \\
\hline
Parameter                &\cite{holmgren1991} &\cite{vaz2007} &\cite{budding2009}& This Work\\
\rule{0pt}{2.2ex}$\mathrm{M_1\,[M_{\odot}]}$  & $4.93\pm 0.05$ & $5.273\pm 0.091$  & $5.13\pm 0.08$ & $5.09\substack{+0.06 \\ -0.05}$\\
\rule{0pt}{2.2ex}$\mathrm{M_2\,[M_{\odot}]}$  & $4.56\pm 0.04$ & $4.783\pm 0.072$  & $4.56\pm 0.07$ & $4.58\substack{+0.05 \\ -0.05}$\\ 
\rule{0pt}{2.2ex}$\mathrm{R_1\,[R_{\odot}]}$  & $3.29\pm 0.06$ & $3.483\pm 0.020$  & $3.41\pm 0.03$ & $3.44\substack{+0.01 \\ -0.01}$\\
\rule{0pt}{2.2ex}$\mathrm{R_2\,[R_{\odot}]}$  & $3.01\pm 0.05$ & $3.109\pm 0.034$  & $3.08\pm 0.03$ & $3.05\substack{+0.01 \\ -0.01}$\\
\rule{0pt}{2.2ex}$\log g_1 \, \mathrm{[dex]}$ & $4.10\pm 0.01$ & $4.068\pm 0.010$  & $4.08\pm 0.01$ & $4.073\substack{+0.004 \\ -0.004}$\\
\rule{0pt}{2.2ex}$\log g_2 \, \mathrm{[dex]}$ & $4.14\pm 0.02$ & $4.128\pm 0.012$  & $4.12\pm 0.01$ & $4.131\substack{+0.004 \\ -0.004}$\\

\hline
\end{tabular}
\tablefoot{
Table compares derived fundamental parameters from this work to previous studies of CW Cep (top) and U Oph (bottom).\\
\tablefoottext{a}{Solution derived using spectroscopic values obtained by \cite{popper1991}} 
\tablefoottext{b}{Solution derived using spectroscopic values obtained by \cite{strickland1992}} }
\end{table*}

\section{Evolutionary Modelling}
\label{section:evol_models}
\subsection{Evolutionary modelling setup}
The updated masses, radii, and effective temperatures of CW\,Cep and U\,Oph provide strong discriminating power against stellar models. As discussed by \cite{constantino2018} and \cite{johnston2019}, however, even such precision does not provide enough of a constraint to uniquely determine the extent of the near-core mixing region. As such, we instead consider the convective core mass, and treat the near-core mixing, parameterised by a diffusive exponentially decaying overshooting prescription with a scaled extent $f_{ov}$, as a nuisance parameter. To do this, we fit each component to a grid of isochrone-clouds as described by \cite{johnston2019}. The isochrone-clouds are constructed from {\sc MESA} tracks computed at solar metallicity $Z=0.014$, with the helium mass fraction fixed according to the Cosmic B-star standard $Y=0.276$ \citep{nieva2012}, with $\alpha_{\rm MLT}=1.8$ using {\sc MESA}-r10398 \citep{mesaI,mesaIV}. We consider an isochrone-cloud at a given age, $\tau$, to cover the range $f_{\rm ov}\in[0.005-0.04;0.005]$ assuming the diffusive exponential description of overshooting implemented in {\sc MESA}. 

We adopt the Mahalanobis distance (MD) as our merit function as applied in \cite{johnston2019} and thorougly discussed in \cite{aerts2018b}. Using the MD as our merit function allows us to account for correlations present amongst model parameters that would otherwise compromise our solution \citep{aerts2018b,johnston2019}. We choose to fit the mass, adopting the errors listed in Table~\ref{tab:derived_pars} instead of interpolating the isochrone-clouds to the dynamical values. Since the MD is a maximum-likelihood point estimator, we perform Monte Carlo simulations (with 10\,000 iterations) to obtain confidence intervals on the model parameters and derived parameters of interest. We select the single best point returned for each iteration. By keeping only the best point, we sample the robustness of our solution given our grid. If we were to keep the best N points, this would instead sample the variance of our solution space as a function of our grid, and observables, which, although is an interesting phenomenon, is ultimately not the focus of this work. After 10\,000 iterations, we bin the resulting distributions for all parameters of interest and apply $95\%$ Highest Posterior Density confidence intervals. The results are listed in Table~\ref{tab:mc_pars}.

\begin{table}
\caption{\label{tab:mc_pars} Monte Carlo Isochrone-cloud modelling $95\%$ confidence intervals for CW\,Cep and U\,Oph.}
\centering
\begin{tabular}{lll}
\hline\hline
Parameter& CW\,Cep & U\,Oph\\
\hline
\rule{0pt}{2.5ex}$\mathrm{Age\,[Myr]}$&$7.0\substack{+1 \\ -1} $        &$57.5\substack{+5.0 \\ -2.5}$\\
\rule{0pt}{2.5ex}$f_{\rm ov,1}$       &$0.025\substack{+0.015 \\ -0.02}$ &$0.025\substack{+0.015 \\ -0.015}$\\
\rule{0pt}{2.5ex}$f_{\rm ov,2}$       &$0.030\substack{+0.01 \\ -0.02}$ &$0.015\substack{+0.015 \\ -0.01}$\\
\rule{0pt}{2.5ex}$M_{\rm 1}\,[{\rm M_{\odot}}]$ &$13.00\substack{+0.1 \\ -0.16}$ &$5.08\substack{+0.07 \\ -0.06}$\\
\rule{0pt}{2.5ex}$M_{\rm 2}\,[{\rm M_{\odot}}]$ &$12.00\substack{+0.11 \\ -0.12}$ &$4.60\substack{+0.05 \\ -0.05}$\\
\rule{0pt}{2.5ex}$X_{\rm c,1}$        &$0.54\substack{+0.01 \\ -0.03}$&$0.48\substack{+0.02 \\ -0.04}$\\
\rule{0pt}{2.5ex}$X_{\rm c,2}$        &$0.57\substack{+0.01 \\ -0.03}$&$0.51\substack{+0.03 \\ -0.02}$\\
\rule{0pt}{2.5ex}$M_{\rm cc,1}\,[{\rm M_{\odot}}]$ &$4.34\substack{+0.11 \\ -0.29}$ &$1.05\substack{+0.08 \\ -0.11}$\\
\rule{0pt}{2.5ex}$M_{\rm cc,2}\,[{\rm M_{\odot}}]$ &$3.86\substack{+0.12 \\ -0.19}$ &$0.93\substack{+0.06 \\ -0.05}$\\
\hline
\end{tabular}

\end{table}

\subsection{Modelling results and discussion}
\label{section:discussion}
The wide range of dynamical solutions for both CW\,Cep and U\,Oph shown in Table~\ref{tab:derived_pars} gives reason for pause. The spread between the minimum and maximum reported solutions is several times larger than the formal uncertainties reported, despite the fact that the same photometric data-sets were used by different studies. The main difference across the individual solutions is the mass ratio, or more fundamentally the spectroscopic data-sets. Furthermore, each set of radial velocities used to calculate the mass ratio was determined using different methods. Most critically, this translates into a large disparity in estimated ages for these systems, and therefore by necessity the estimated  internal mixing. This is easily seen in the spread in ages for each system discussed earlier in Section~\ref{section:overview}. In addition to different masses and radii being used, different effective temperatures are also fit in the individual modelling efforts. In the end, these differences effectively mean that each study is modelling a different system. This highlights the need for a systematic evaluation of the accuracy versus the precision of dynamical and spectroscopic solutions for well studied eclipsing binaries. However, that is beyond the scope of the work in this manuscript. We note that future studies which entail modelling efforts of samples comprised of systems which were not homogeneously analysed must consider the systematic differences between different methods. We also note the necessity in allowing the mass ratio, $q$, to vary. In the case where the mass ratio is fixed, the dynamical solution returns artificially high precision to the fourth or later decimal place. Given the high-precision {\' e}chelle spectra, combined with state-of-the-art {\sc spd} and MCMC methodologies, we find our solution to be more robust than previous solutions. As such, for the remainder of the discussion, we only consider the results obtained in this work.

As discussed previously, the evolutionary modelling of eclipsing binaries involves several parameter degeneracies. While many studies attempt to constrain near core mixing, the modelling procedure is not directly sensitive to the details of the prescriptions of these phenomena, but rather to their consequences. As such, any inference drawn on stellar rotation, convective overshooting, and/or magnetism from evolutionary modelling is convoluted with additional effects and uncertainties, at least some of which can be attributed to the implementation of such effects as diffusive processes in stellar structure and evolution codes. Due to this, although we have shown that CW~Cep A and B and U~Oph A and B are rotating at roughly a quarter of their critical rotation rates, any internal mixing caused by this will be degenerate with mixing caused by convective overshooting. Therefore, reflecting this and the discussions presented by \cite{constantino2018} and \cite{johnston2019}, we gear our discussion towards the core properties, which evolutionary modelling is more directly sensitive to since these properties dictate the stellar evolutionary sequence.  

\begin{figure*}
\centering
\includegraphics[width=\linewidth]{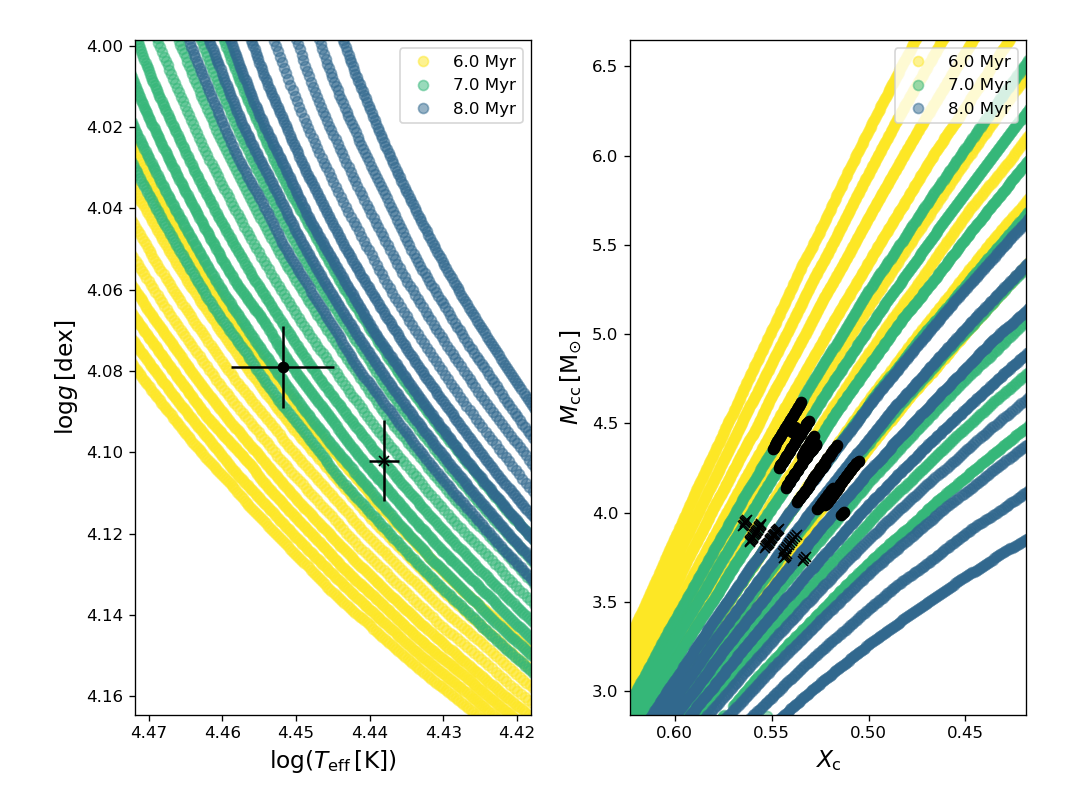}
\caption{{\bf Left}: Isochrone-clouds for the ages reported in Table~\ref{tab:mc_pars} with the spectroscopic uncertainties plotted for CW\,Cep A and B in black. {\bf Right}: Convective-mass plotted against the core hydrogen content for the isochrone-clouds shown in the left panel in grey. Those regions which are allowed by the spectroscopic uncertainties for CW\,Cep A and B are shown as black circles and x's, respectively.}
\label{fig:cw_cep_kiel}
\end{figure*}

\begin{figure*}
\centering
\includegraphics[width=\linewidth]{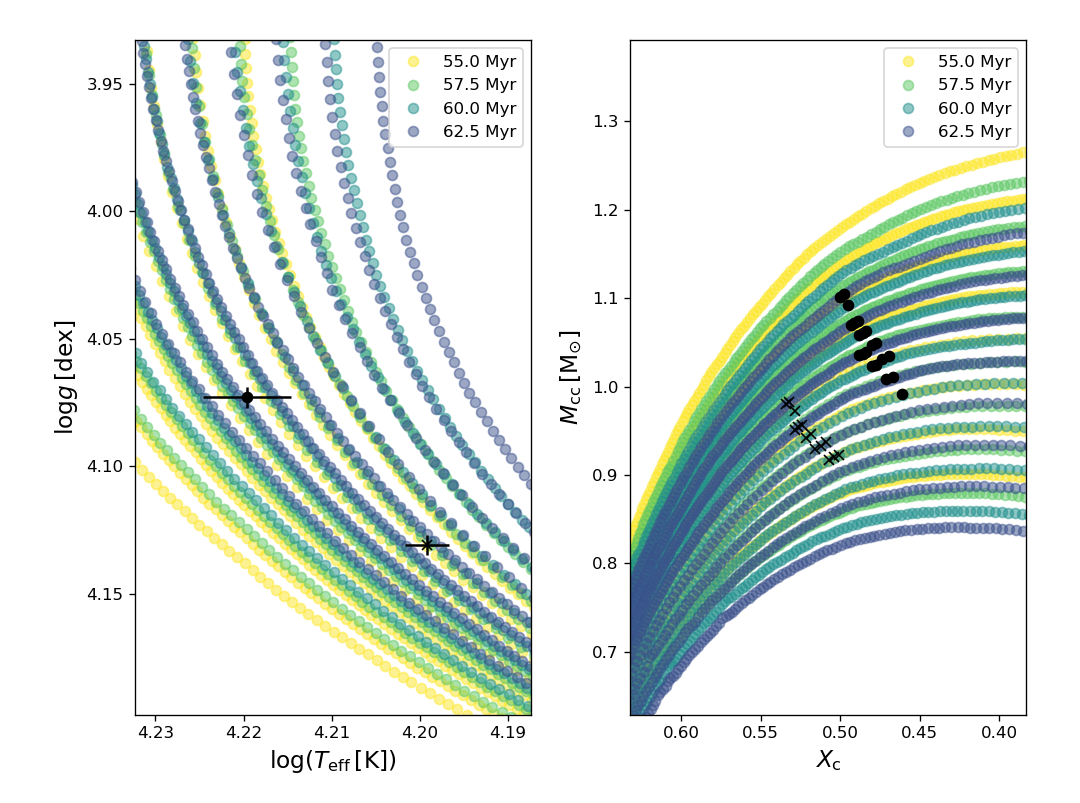}
\caption{Same as Fig.~\ref{fig:cw_cep_kiel}, but for U\,Oph.}
\label{fig:u_oph_kiel}
\end{figure*}

Despite the per-cent level precision on dynamical quantities provided by the binary solution, our modelling  could not provide a constrained range for the extent of near-core mixing for the primary of CW\,Cep or for either component in U\,Oph. However, our modelling shows that CW\,Cep\,B requires a large amount of internal mixing to have its current observed properties and be co-evolutionary with CW\,Cep\,A. Stated differently, CW\,Cep\,B requires a more massive core than models based solely on the Schwarzchild criterion, otherwise it would appear as a different age compared to CW\,Cep\,A. The left panels of Figures \ref{fig:cw_cep_kiel} and \ref{fig:u_oph_kiel} show how the isochrone-clouds of ages reported in Table\,\ref{tab:mc_pars} cover large, and often overlapping, regions of the spectroscopic parameter space due to the spread in near-core mixing. As can be seen in the accompanying right panels of said figures, this translates to a generally more confined region in core properties shown in black circles and black x's for the primary and secondary, respectively for either system. At the age and mass range for CW\,Cep, the components have not progressed sufficiently through their MS lifetimes to be able to critically constrain their core properties, with the primary and secondary being $\sim27\%$ and $\sim21\%$ through their MS lifetimes, respectively. The cores of CW\,Cep\,A and B contain $\sim33\substack{+1\\-4}\%$ and $\sim32\substack{+1\\-2}\%$ of the total mass, respectively.As for U\,Oph, the primary and secondary are $\sim40\%$ and $\sim31\%$ through their respective MS lifetimes. In this case, the resulting core parameter regions are more constrained. The cores of U\,Oph\,A and B contain $\sim19\substack{+3\\-1}\%$ and $\sim19\substack{+2\\-0.5}\%$ of the total mass, respectively.

The age estimate we find for CW\,Cep largely agrees with the estimates of \cite{jordi1996} and \cite{clausen_gimenez_1991}, but is nearly twice as old as the solution reported by \cite{ribas2000b}. Our age estimate for U\,Oph is considerably higher than the median reported value in the literature, but does agree with the upper limits of those solutions reported with a lower metallicity closer to the value we find and use in our modelling. We note that in both cases, the estimated ages agree with those reported by previous studies. However, the solutions presented here are systematically older than those presented by \cite{schneider2014} for both CW\,Cep and U\,Oph. Furthermore, our solution for CW\,Cep show the components as much less progressed along the MS ($\sim27\%$ and $\sim21\%$) than compared to \cite{schneider2014} ($\sim35-40\%$ and $\sim30-35\%$). This discrepancy is similar in the case of U\,Oph, with \cite{schneider2014} reported U\,Oph A and B being $50\%$ and $40\%$ through the MS, respectively, compared to the $\sim40\%$ and $\sim31\%$ progress that we report. However, \cite{schneider2014} use a solution which is $\sim5\%$ more massive compared to ours. This again highlights the need for a homogeneously analysed sample to draw inference on trends in core properties and the physical processes which influence them.

\subsection{Rotation, Synchronicity, and Circularisation}
According to our MCMC estimates, CW~Cep A and U~Oph B are both rotating super-synchronously by $2\sigma$, while CW~Cep B and U~Oph A are both rotating synchronously within uncertainties. By combining the synchronicity parameters $f_1,2$ and the derived parameters for each binary, we can calculate the un-projected rotational velocity of each component. We could perform this calculation with the projected rotation velocities obtained in Section~\ref{section:spectra_atmosphere}, however, since we applied this information as a prior in our MCMC analysis, using the synchronicity parameters makes use of the same information. From this, we find that CW~Cep A and B are rotating at $v_A=107\pm3\,\,{\rm km\,s^{-1}}$ and $v_B=97\pm3\,\,{\rm km\,s^{-1}}$, respectively, and U~Oph A and B are rotating at $v_A=111\pm7\,\,{\rm km\,s^{-1}}$ and $v_B=107\pm6\,\,{\rm km\,s^{-1}}$, respectively. 

We also calculate the critical rotation rates for each star using the parameters listed in Table~\ref{tab:derived_pars}. We find the critical rotation rates for CW~Cep A and B to be $v_{\rm crit, A}=551\pm3\,\,{\rm km\,s^{-1}}$ and $v_{\rm crit, B}=546\pm3\,\,{\rm km\,s^{-1}}$, respectively. This reveals that CW~Cep A and B are rotating at $19.6\pm0.6\%$ and $17.7\pm0.6\%$ their critical rates, respectively. For U~Oph A and B, we find $v_{\rm crit, A}=434\pm3\,\,{\rm km\,s^{-1}}$ and $v_{\rm crit, B}=437\pm2\,\,{\rm km\,s^{-1}}$, respectively. Thus revealing them to be rotating at $25.6\pm2\%$ and $24.4\pm1\%$ their critical rates, respectively. 

Tidal theory gives predictions of synchronisation and circularisation timescales for binary systems \citep{zahn1975,zahn1977}. Using the results of our modelling, we find $\tau_{sync}=0.606\pm0.002$ Myr and $\tau_{circ}=62.9\pm1$ Myr for CW~Cep and $\tau_{sync}=0.0879\pm0.0005$ Myr and $\tau_{circ}=4.63\pm0.02$ Myr for U~Oph. The ages obtained in our isochrone-cloud modelling are an order of magnitude larger than the theoretical synchronisation timescales for either CW~Cep or U~Oph. CW~Cep is significantly younger than its theoretical circularisation timescale, and given the masses of the components, it will evolve beyond the MS before circularisation occurs. U~Oph, however, is already significantly older than its circularisation timescale, but is still observed to be eccentric. This observation fits with the presence of a third body that likely sends the system through Kozai-Lidov cycles, as opposed to having a constantly decaying eccentricity.

\section{Conclusions}
\label{section:conclusions}
Contemporary binary modelling techniques have the ability to provide per-cent level (or better) precision on fundamental stellar parameter estimates to be compared against evolutionary models. These parameter estimates have been used by numerous studies, including this one, to attempt to constrain poorly understood near-core mixing processes which cause deviations from nominal stellar evolution. However, no clear consensus exists in the literature as to wether or not this is possible. 

In this work we obtained and analysed new spectroscopic observations on the intermediate- to high-mass binaries CW\,Cep and U\,Oph. Our analysis yielded an updated mass-ratio to be used for lightcurve modelling, as well as the first abundance patterns for these systems. The abundance patterns were revealed to be roughly solar, which was exploited in the isochrone-cloud evolutionary modelling. 

We performed lightcurve modelling using a Bayesian MCMC optimization routine wrapped around the PHOEBE binary modelling code to obtain updated and highly precise mass and radius estimates. These estimates roughly agree with past studies, but the spread in reported solutions is much larger than the precision reported for any solution. This raises a concern in relation to the robustness of the accuracy of a solution versus its precision. To test the consistency of our solutions, we compared the luminosities from binary modelling with those calculated from Gaia parallaxes. Furthermore, given the close separation of the components in these systems, they are ideal candidates for investigating the influence of the inclusion of second order physics such as Doppler boosting and reflection on resulting modelled core properties. However, such analysis requires high-precision space photometry which has yet to be assembled for these systems, but will be done soon by the {\it TESS} mission \citep{ricker2015}. Additionally, the high precision orbital and dynamical solutions allowed us to investigate the rotation rates and tidal synchronisation and circularisation timescales for both systems.

Using our updated dynamical solution and temperatures, we performed isochrone-cloud modelling following the procedure as described by \cite{johnston2019} to obtain estimates on both model input parameters, as well as derived parameters such as the core properties. Our results reveal that, given model degeneracies, we cannot critically constrain the extent of near-core mixing. We do, however, constrain the core mass and hydrogen content for both components of CW\,Cep and U\,Oph, as these quantities directly dictate the current evolutionary status of a star. We compare our results to those of \cite{schneider2014}, who performed a similar analysis, but which assumed rotational mixing instead of exponentially decaying diffusive convetive overshooting in their evolutionary models. Combined with the alarming spread in reported dynamical solutions shown in Table~\ref{tab:derived_pars}, our comparison highlights the need for a homogeneously analysed sample to be able to make meaningful inference on internal physical processes such as convective overshooting, rotational and pulsational mixing, and magnetism \citep{aerts2014}. 

Finally, we ask that future studies which perform evolutionary modelling report core masses of their solutions in addition to the overshooting extent or near-core rotation rate. 

\section*{Acknowledgements}
We thank the referee for their helpful comments which have improved the manuscript. We thank Dr. Dominic Bowman and Prof. Hugues Sana for their discussions pertaining to error determination. The research leading to these results has received funding from the European Research Council (ERC) under the European Union’s Horizon 2020 research and innovation programme (grant agreement N$^\circ$670519: MAMSIE ; CJ \& AT), from the Research Foundation Flanders (FWO) under grant agreement G0A2917N (BlackGEM ; CJ), from the KU\,Leuven Research Council (grant C16/18/005: PARADISE ; AT), from the Research Foundation Flanders (FWO) under grant agreement G0H5416N (ERC Runner Up Project ; AT), as well as from the BELgian federal Science Policy Office (BELSPO) through PRODEX grant PLATO (AT). KP acknowledges financial support from the Croatian Science Foundation under grant IP-2014-09-8656 (STARDUST). The computational resources and services used in this work were provided by the VSC (Flemish Supercomputer Center), funded by the Research Foundation - Flanders (FWO) and the Flemish Government – department EWI. Based on observations made with the Mercator Telescope, operated on the island of La Palma by the Flemmish Community, at the Spanish Observatorio del Roque de los Muchachos of the Instituto de Astrofísica de Canarias.

\bibliographystyle{aa}
\bibliography{johnston_accepted_v1.bib}

\begin{thebibliography}{131}
\expandafter\ifx\csname natexlab\endcsname\relax\def\natexlab#1{#1}\fi

\bibitem[{{Aerts} \& {Harmanec}(2004)}]{aerts2004}
{Aerts}, C. \& {Harmanec}, P. 2004, in Astronomical Society of the Pacific
  Conference Series, Vol. 318, Spectroscopically and Spatially Resolving the
  Components of the Close Binary Stars, ed. R.~W. {Hilditch}, H.~{Hensberge},
  \& K.~{Pavlovski}, 325--333

\bibitem[{{Aerts} {et~al.}(2019){Aerts}, {Mathis}, \& {Rogers}}]{aerts2019}
{Aerts}, C., {Mathis}, S., \& {Rogers}, T. 2019, ARA\&A, arXiv:1809.07779

\bibitem[{{Aerts} {et~al.}(2014){Aerts}, {Molenberghs}, {Kenward}, \&
  {Neiner}}]{aerts2014}
{Aerts}, C., {Molenberghs}, G., {Kenward}, M.~G., \& {Neiner}, C. 2014, \apj,
  781, 88

\bibitem[{{Aerts} {et~al.}(2018){Aerts}, {Molenberghs}, {Michielsen},
  {Pedersen}, {Bj{\"o}rklund}, {Johnston}, {Mombarg}, {Bowman}, {Buysschaert},
  {P{\'a}pics}, {Sekaran}, {Sundqvist}, {Tkachenko}, {Truyaert}, {Van Reeth},
  \& {Vermeyen}}]{aerts2018b}
{Aerts}, C., {Molenberghs}, G., {Michielsen}, M., {et~al.} 2018, \apjs, 237, 15

\bibitem[{{Andersen} {et~al.}(1990){Andersen}, {Nordstroem}, \&
  {Clausen}}]{andersen1990}
{Andersen}, J., {Nordstroem}, B., \& {Clausen}, J.~V. 1990, \apj, 363, L33

\bibitem[{{Asplund} {et~al.}(2009){Asplund}, {Grevesse}, {Sauval}, \&
  {Scott}}]{asplund+2009}
{Asplund}, M., {Grevesse}, N., {Sauval}, A.~J., \& {Scott}, P. 2009, ARA\&A,
  47, 481

\bibitem[{{Beck} {et~al.}(2018{\natexlab{a}}){Beck}, {Kallinger}, {Pavlovski},
  {Palacios}, {Tkachenko}, {Mathis}, {Garc{\'\i}a}, {Corsaro}, {Johnston},
  {Mosser}, {Ceillier}, {do Nascimento}, \& {Raskin}}]{beck2018a}
{Beck}, P.~G., {Kallinger}, T., {Pavlovski}, K., {et~al.} 2018{\natexlab{a}},
  \aap, 612, A22

\bibitem[{{Beck} {et~al.}(2018{\natexlab{b}}){Beck}, {Mathis}, {Gallet},
  {Charbonnel}, {Benbakoura}, {Garc{\'\i}a}, \& {do Nascimento}}]{beck2018b}
{Beck}, P.~G., {Mathis}, S., {Gallet}, F., {et~al.} 2018{\natexlab{b}}, \mnras,
  479, L123

\bibitem[{{Blaauw}(1961)}]{blaauw1961}
{Blaauw}, A. 1961, Bulletin of the Astronomical Institutes of the Netherlands,
  15, 265

\bibitem[{{Blaauw} {et~al.}(1959){Blaauw}, {Hiltner}, \&
  {Johnson}}]{blaauw1959}
{Blaauw}, A., {Hiltner}, W.~A., \& {Johnson}, H.~L. 1959, \apj, 130, 69

\bibitem[{{Bonanos} {et~al.}(2006){Bonanos}, {Stanek}, {Kudritzki}, {Macri},
  {Sasselov}, {Kaluzny}, {Stetson}, {Bersier}, {Bresolin}, {Matheson},
  {Mochejska}, {Przybilla}, {Szentgyorgyi}, {Tonry}, \& {Torres}}]{bonanos2006}
{Bonanos}, A.~Z., {Stanek}, K.~Z., {Kudritzki}, R.~P., {et~al.} 2006, \apj,
  652, 313

\bibitem[{{Briquet} {et~al.}(2007){Briquet}, {Morel}, {Thoul}, {Scuflaire},
  {Miglio}, {Montalb{\'a}n}, {Dupret}, \& {Aerts}}]{briquet2007}
{Briquet}, M., {Morel}, T., {Thoul}, A., {et~al.} 2007, \mnras, 381, 1482

\bibitem[{{Brott} {et~al.}(2011{\natexlab{a}}){Brott}, {de Mink}, {Cantiello},
  {Langer}, {de Koter}, {Evans}, {Hunter}, {Trundle}, \& {Vink}}]{brott2011a}
{Brott}, I., {de Mink}, S.~E., {Cantiello}, M., {et~al.} 2011{\natexlab{a}},
  \aap, 530, A115

\bibitem[{{Brott} {et~al.}(2011{\natexlab{b}}){Brott}, {Evans}, {Hunter}, {de
  Koter}, {Langer}, {Dufton}, {Cantiello}, {Trundle}, {Lennon}, {de Mink},
  {Yoon}, \& {Anders}}]{brott2011b}
{Brott}, I., {Evans}, C.~J., {Hunter}, I., {et~al.} 2011{\natexlab{b}}, \aap,
  530, A116

\bibitem[{{Budding} {et~al.}(2009){Budding}, {Inlek}, \&
  {Demircan}}]{budding2009}
{Budding}, E., {Inlek}, G., \& {Demircan}, O. 2009, \mnras, 393, 501

\bibitem[{{Butler} {et~al.}(1984){Butler}, {Mendoza}, \&
  {Zeippen}}]{butler1984}
{Butler}, K., {Mendoza}, C., \& {Zeippen}, C.~J. 1984, Journal of Physics B
  Atomic Molecular Physics, 17, 2039

\bibitem[{{Charbonneau}(1995)}]{charbonneau1995}
{Charbonneau}, P. 1995, \apjs, 101, 309

\bibitem[{{Claret}(2007)}]{claret2007}
{Claret}, A. 2007, \aap, 475, 1019

\bibitem[{{Claret} \& {Torres}(2016)}]{claret2016}
{Claret}, A. \& {Torres}, G. 2016, \aap, 592, A15

\bibitem[{{Claret} \& {Torres}(2017)}]{claret2017}
{Claret}, A. \& {Torres}, G. 2017, \apj, 849, 18

\bibitem[{{Claret} \& {Torres}(2018)}]{claret2018}
{Claret}, A. \& {Torres}, G. 2018, \apj, 859, 100

\bibitem[{{Claret} \& {Torres}(2019)}]{claret2019}
{Claret}, A. \& {Torres}, G. 2019, arXiv e-prints, arXiv:1904.02714

\bibitem[{{Clausen} \& {Gimenez}(1991)}]{clausen_gimenez_1991}
{Clausen}, J.~V. \& {Gimenez}, A. 1991, \aap, 241, 98

\bibitem[{{Clements} \& {Neff}(1979)}]{clements1979}
{Clements}, G.~L. \& {Neff}, J.~S. 1979, \apjs, 41, 1

\bibitem[{{Constantino} \& {Baraffe}(2018)}]{constantino2018}
{Constantino}, T. \& {Baraffe}, I. 2018, \aap, 618, A177

\bibitem[{{De Cat} {et~al.}(2000){De Cat}, {Aerts}, {De Ridder}, {Kolenberg},
  {Meeus}, \& {Decin}}]{decat2000}
{De Cat}, P., {Aerts}, C., {De Ridder}, J., {et~al.} 2000, \aap, 355, 1015

\bibitem[{{De Cat} {et~al.}(2004){De Cat}, {De Ridder}, {Hensberge}, \&
  {Ilijic}}]{decat2004}
{De Cat}, P., {De Ridder}, J., {Hensberge}, H., \& {Ilijic}, S. 2004, in
  Astronomical Society of the Pacific Conference Series, Vol. 318,
  Spectroscopically and Spatially Resolving the Components of the Close Binary
  Stars, ed. R.~W. {Hilditch}, H.~{Hensberge}, \& K.~{Pavlovski}, 338--341

\bibitem[{{de Mink} {et~al.}(2013){de Mink}, {Langer}, {Izzard}, {Sana}, \& {de
  Koter}}]{deMink2013}
{de Mink}, S.~E., {Langer}, N., {Izzard}, R.~G., {Sana}, H., \& {de Koter}, A.
  2013, \apj, 764, 166

\bibitem[{{Eaton} \& {Ward}(1973)}]{eaton1973}
{Eaton}, J.~A. \& {Ward}, D.~H. 1973, \apj, 185, 921

\bibitem[{{Ekstr{\"o}m} {et~al.}(2012){Ekstr{\"o}m}, {Georgy}, {Eggenberger},
  {Meynet}, {Mowlavi}, {Wyttenbach}, {Granada}, {Decressin}, {Hirschi},
  {Frischknecht}, {Charbonnel}, \& {Maeder}}]{ekstrom2012}
{Ekstr{\"o}m}, S., {Georgy}, C., {Eggenberger}, P., {et~al.} 2012, \aap, 537,
  A146

\bibitem[{{Ekstr{\"o}m} {et~al.}(2018){Ekstr{\"o}m}, {Meynet}, {Georgy}, \&
  {Granada}}]{ekstrom2018}
{Ekstr{\"o}m}, S., {Meynet}, G., {Georgy}, C., \& {Granada}, A. 2018, Memorie
  della Societa Astronomica Italiana, 89, 50

\bibitem[{{Elias} {et~al.}(2008){Elias}, {Koch}, \& {Pfeiffer}}]{elias2008}
{Elias}, N.~M., I., {Koch}, R.~H., \& {Pfeiffer}, R.~J. 2008, \aap, 489, 911

\bibitem[{{Erdem} {et~al.}(2004){Erdem}, {Soydugan}, {Soydugan}, {{\"O}zdemir},
  {{\c{C}}i{\c{c}}ek}, {Bulut}, {Demircan}, \& {Budding}}]{erdem2004}
{Erdem}, A., {Soydugan}, E., {Soydugan}, F., {et~al.} 2004, Astron. Nachr.,
  325, 336

\bibitem[{{Foreman-Mackey} {et~al.}(2013){Foreman-Mackey}, {Hogg}, {Lang}, \&
  {Goodman}}]{foremanMackey2013}
{Foreman-Mackey}, D., {Hogg}, D.~W., {Lang}, D., \& {Goodman}, J. 2013, \pasp,
  125, 306

\bibitem[{{Frieboes-Conde} \& {Herczeg}(1973)}]{frieboes1973}
{Frieboes-Conde}, H. \& {Herczeg}, T. 1973, \aaps, 12, 1

\bibitem[{{Gallenne} {et~al.}(2016){Gallenne}, {Pietrzy{\'n}ski}, {Graczyk},
  {Konorski}, {Kervella}, {M{\'e}rand}, {Gieren}, {Anderson}, \&
  {Villanova}}]{gallenne2016}
{Gallenne}, A., {Pietrzy{\'n}ski}, G., {Graczyk}, D., {et~al.} 2016, \aap, 586,
  A35

\bibitem[{{Giddings}(1981)}]{giddings1981}
{Giddings}, J.~R. 1981, PhD thesis, -

\bibitem[{{Gimenez} {et~al.}(1987){Gimenez}, {Kim}, \& {Nha}}]{gimenez1987}
{Gimenez}, A., {Kim}, C.-H., \& {Nha}, I.-S. 1987, \mnras, 224, 543

\bibitem[{{Gimenez} {et~al.}(1990){Gimenez}, {Rolland}, {Garcia}, \&
  {Clausen}}]{gimenez1990}
{Gimenez}, A., {Rolland}, A., {Garcia}, J.~M., \& {Clausen}, J.~V. 1990, \aaps,
  86, 259

\bibitem[{{Grunhut} {et~al.}(2013){Grunhut}, {Wade}, {Leutenegger}, {Petit},
  {Rauw}, {Neiner}, {Martins}, {Cohen}, {Gagn{\'e}}, {Ignace}, {Mathis}, {de
  Mink}, {Moffat}, {Owocki}, {Shultz}, {Sundqvist}, \& {MiMeS
  Collaboration}}]{grunhut2013}
{Grunhut}, J.~H., {Wade}, G.~A., {Leutenegger}, M., {et~al.} 2013, \mnras, 428,
  1686

\bibitem[{{Guinan} {et~al.}(1998){Guinan}, {Fitzpatrick}, {DeWarf}, {Maloney},
  {Maurone}, {Ribas}, {Pritchard}, {Bradstreet}, \& {Gim{\'e}nez}}]{guinan1998}
{Guinan}, E.~F., {Fitzpatrick}, E.~L., {DeWarf}, L.~E., {et~al.} 1998, \apj,
  509, L21

\bibitem[{{Guinan} {et~al.}(2000){Guinan}, {Ribas}, {Fitzpatrick},
  {Gim{\'e}nez}, {Jordi}, {McCook}, \& {Popper}}]{guinan2000}
{Guinan}, E.~F., {Ribas}, I., {Fitzpatrick}, E.~L., {et~al.} 2000, \apj, 544,
  409

\bibitem[{{Hadrava}(1995)}]{hadrava1995}
{Hadrava}, P. 1995, \aaps, 114, 393

\bibitem[{{Hambleton} {et~al.}(2016){Hambleton}, {Kurtz}, {Pr{\v{s}}a},
  {Quinn}, {Fuller}, {Murphy}, {Thompson}, {Latham}, \&
  {Shporer}}]{hambleton2016}
{Hambleton}, K., {Kurtz}, D.~W., {Pr{\v{s}}a}, A., {et~al.} 2016, \mnras, 463,
  1199

\bibitem[{{Han} {et~al.}(2002){Han}, {Kim}, {Lee}, \& {Koch}}]{han2002}
{Han}, W., {Kim}, C.-H., {Lee}, W.-B., \& {Koch}, R.~H. 2002, \aj, 123, 2724

\bibitem[{{Hensberge} \& {Pavlovski}(2007)}]{hensberge2007}
{Hensberge}, H. \& {Pavlovski}, K. 2007, in IAU Symposium, Vol. 240, Binary
  Stars as Critical Tools \&amp; Tests in Contemporary Astrophysics, ed. W.~I.
  {Hartkopf}, P.~{Harmanec}, \& E.~F. {Guinan}, 136--147

\bibitem[{{Hensberge} {et~al.}(2000){Hensberge}, {Pavlovski}, \&
  {Verschueren}}]{hensberge2000}
{Hensberge}, H., {Pavlovski}, K., \& {Verschueren}, W. 2000, \aap, 358, 553

\bibitem[{{Herrero} {et~al.}(1992){Herrero}, {Kudritzki}, {Vilchez}, {Kunze},
  {Butler}, \& {Haser}}]{herrero1992}
{Herrero}, A., {Kudritzki}, R.~P., {Vilchez}, J.~M., {et~al.} 1992, \aap, 261,
  209

\bibitem[{{Higl} \& {Weiss}(2017)}]{higl2017}
{Higl}, J. \& {Weiss}, A. 2017, \aap, 608, A62

\bibitem[{{Hirschi} {et~al.}(2014){Hirschi}, {den Hartogh}, {Cristini},
  {Georgy}, \& {Pignatari}}]{hirschi2014}
{Hirschi}, R., {den Hartogh}, J., {Cristini}, A., {Georgy}, C., \& {Pignatari},
  M. 2014, in XIII Nuclei in the Cosmos (NIC XIII), 1

\bibitem[{{Holmgren} {et~al.}(1991){Holmgren}, {Hill}, \&
  {Fisher}}]{holmgren1991}
{Holmgren}, D.~E., {Hill}, G., \& {Fisher}, W. 1991, \aap, 248, 129

\bibitem[{{Huffer} \& {Kopal}(1951)}]{huffer1951}
{Huffer}, C.~M. \& {Kopal}, Z. 1951, \apj, 114, 297

\bibitem[{{Ignace} {et~al.}(2017){Ignace}, {Hole}, {Oskinova}, \&
  {Rotter}}]{ignace2017}
{Ignace}, R., {Hole}, K.~T., {Oskinova}, L.~M., \& {Rotter}, J.~P. 2017, \apj,
  850, 82

\bibitem[{{Ilijic} {et~al.}(2004){Ilijic}, {Hensberge}, {Pavlovski}, \&
  {Freyhammer}}]{ilijic2004}
{Ilijic}, S., {Hensberge}, H., {Pavlovski}, K., \& {Freyhammer}, L.~M. 2004, in
  Astronomical Society of the Pacific Conference Series, Vol. 318,
  Spectroscopically and Spatially Resolving the Components of the Close Binary
  Stars, ed. R.~W. {Hilditch}, H.~{Hensberge}, \& K.~{Pavlovski}, 111--113

\bibitem[{{Ivezi{\'c}} {et~al.}(2014){Ivezi{\'c}}, {Connelly}, {VanderPlas}, \&
  {Gray}}]{ivezic2014}
{Ivezi{\'c}}, {\v{Z}}., {Connelly}, A.~J., {VanderPlas}, J.~T., \& {Gray}, A.
  2014, {Statistics, Data Mining, and Machine Learningin Astronomy}

\bibitem[{{Johnston} {et~al.}(2017){Johnston}, {Buysschaert}, {Tkachenko},
  {Aerts}, \& {Neiner}}]{johnston2017}
{Johnston}, C., {Buysschaert}, B., {Tkachenko}, A., {Aerts}, C., \& {Neiner},
  C. 2017, \mnras, 469, L118

\bibitem[{{Johnston} {et~al.}(2019){Johnston}, {Tkachenko}, {Aerts},
  {Molenberghs}, {Bowman}, {Pedersen}, {Buysschaert}, \&
  {P{\'a}pics}}]{johnston2019}
{Johnston}, C., {Tkachenko}, A., {Aerts}, C., {et~al.} 2019, \mnras, 482, 1231

\bibitem[{{Jordi} {et~al.}(1996){Jordi}, {Trullols}, \&
  {Galadi-Enriquez}}]{jordi1996}
{Jordi}, C., {Trullols}, E., \& {Galadi-Enriquez}, D. 1996, \aap, 312, 499

\bibitem[{{Kaemper}(1986)}]{kaemper1986}
{Kaemper}, B.~C. 1986, \apss, 120, 167

\bibitem[{{Kirkby-Kent} {et~al.}(2018){Kirkby-Kent}, {Maxted}, {Serenelli},
  {Turner}, {Evans}, {Anderson}, {Hellier}, \& {West}}]{kirkbyKent2018}
{Kirkby-Kent}, J.~A., {Maxted}, P.~F.~L., {Serenelli}, A.~M., {et~al.} 2018,
  \aap, 620, C5

\bibitem[{{Koch} \& {Koegler}(1977)}]{koch1977}
{Koch}, R.~H. \& {Koegler}, C.~A. 1977, \apj, 214, 423

\bibitem[{{Kochukhov} {et~al.}(2018){Kochukhov}, {Johnston}, {Alecian}, \&
  {Wade}}]{kochukhov2018}
{Kochukhov}, O., {Johnston}, C., {Alecian}, E., \& {Wade}, G.~A. 2018, \mnras,
  478, 1749

\bibitem[{{Kolbas} {et~al.}(2014){Kolbas}, {Dervișo{\u{g}}lu}, {Pavlovski}, \&
  {Southworth}}]{kolbas2014}
{Kolbas}, V., {Dervișo{\u{g}}lu}, A., {Pavlovski}, K., \& {Southworth}, J.
  2014, \mnras, 444, 3118

\bibitem[{{Lindegren} {et~al.}(2018){Lindegren}, {Hern{\'a}ndez}, {Bombrun},
  {Klioner}, {Bastian}, {Ramos-Lerate}, {de Torres}, {Steidelm{\"u}ller},
  {Stephenson}, {Hobbs}, {Lammers}, {Biermann}, {Geyer}, {Hilger}, {Michalik},
  {Stampa}, {McMillan}, {Casta{\~n}eda}, {Clotet}, {Comoretto}, {Davidson},
  {Fabricius}, {Gracia}, {Hambly}, {Hutton}, {Mora}, {Portell}, {van Leeuwen},
  {Abbas}, {Abreu}, {Altmann}, {Andrei}, {Anglada}, {Balaguer-N{\'u}{\~n}ez},
  {Barache}, {Becciani}, {Bertone}, {Bianchi}, {Bouquillon}, {Bourda},
  {Br{\"u}semeister}, {Bucciarelli}, {Busonero}, {Buzzi}, {Cancelliere},
  {Carlucci}, {Charlot}, {Cheek}, {Crosta}, {Crowley}, {de Bruijne}, {de
  Felice}, {Drimmel}, {Esquej}, {Fienga}, {Fraile}, {Gai}, {Garralda},
  {Gonz{\'a}lez-Vidal}, {Guerra}, {Hauser}, {Hofmann}, {Holl}, {Jordan},
  {Lattanzi}, {Lenhardt}, {Liao}, {Licata}, {Lister}, {L{\"o}ffler},
  {Marchant}, {Martin-Fleitas}, {Messineo}, {Mignard}, {Morbidelli}, {Poggio},
  {Riva}, {Rowell}, {Salguero}, {Sarasso}, {Sciacca}, {Siddiqui}, {Smart},
  {Spagna}, {Steele}, {Taris}, {Torra}, {van Elteren}, {van Reeven}, \&
  {Vecchiato}}]{lindegren2018}
{Lindegren}, L., {Hern{\'a}ndez}, J., {Bombrun}, A., {et~al.} 2018, \aap, 616,
  A2

\bibitem[{{Luri} {et~al.}(2018){Luri}, {Brown}, {Sarro}, {Arenou},
  {Bailer-Jones}, {Castro-Ginard}, {de Bruijne}, {Prusti}, {Babusiaux}, \&
  {Delgado}}]{luri2018}
{Luri}, X., {Brown}, A.~G.~A., {Sarro}, L.~M., {et~al.} 2018, \aap, 616, A9

\bibitem[{{Lyubimkov} {et~al.}(2005){Lyubimkov}, {Rostopchin}, {Rachkovskaya},
  {Poklad}, \& {Lambert}}]{lyubimkov2005}
{Lyubimkov}, L.~S., {Rostopchin}, S.~I., {Rachkovskaya}, T.~M., {Poklad},
  D.~B., \& {Lambert}, D.~L. 2005, \mnras, 358, 193

\bibitem[{{Maeder}(2009)}]{maeder2009}
{Maeder}, A. 2009, {Physics, Formation and Evolution of Rotating Stars}

\bibitem[{{Moravveji} {et~al.}(2015){Moravveji}, {Aerts}, {P{\'a}pics},
  {Triana}, \& {Vandoren}}]{moravveji2015}
{Moravveji}, E., {Aerts}, C., {P{\'a}pics}, P.~I., {Triana}, S.~A., \&
  {Vandoren}, B. 2015, \aap, 580, A27

\bibitem[{{Moravveji} {et~al.}(2016){Moravveji}, {Townsend}, {Aerts}, \&
  {Mathis}}]{moravveji2016}
{Moravveji}, E., {Townsend}, R.~H.~D., {Aerts}, C., \& {Mathis}, S. 2016, \apj,
  823, 130

\bibitem[{{Nha}(1975)}]{nha1975}
{Nha}, I.~S. 1975, \aj, 80, 232

\bibitem[{{Nieva} \& {Przybilla}(2012)}]{nieva2012}
{Nieva}, M.-F. \& {Przybilla}, N. 2012, \aap, 539, A143

\bibitem[{{Pablo} {et~al.}(2017){Pablo}, {Richardson}, {Fuller}, {Rowe},
  {Moffat}, {Kuschnig}, {Popowicz}, {Handler}, {Neiner}, {Pigulski}, {Wade},
  {Weiss}, {Buysschaert}, {Ramiaramanantsoa}, {Bratcher}, {Gerhartz}, {Greco},
  {Hardegree-Ullman}, {Lembryk}, \& {Oswald}}]{pablo2017}
{Pablo}, H., {Richardson}, N.~D., {Fuller}, J., {et~al.} 2017, \mnras, 467,
  2494

\bibitem[{{Pablo} {et~al.}(2015){Pablo}, {Richardson}, {Moffat}, {Corcoran},
  {Shenar}, {Benvenuto}, {Fuller}, {Naz{\'e}}, {Hoffman}, {Miroshnichenko},
  {Ma{\'\i}z Apell{\'a}niz}, {Evans}, {Eversberg}, {Gayley}, {Gull},
  {Hamaguchi}, {Hamann}, {Henrichs}, {Hole}, {Ignace}, {Iping}, {Lauer},
  {Leutenegger}, {Lomax}, {Nichols}, {Oskinova}, {Owocki}, {Pollock},
  {Russell}, {Waldron}, {Buil}, {Garrel}, {Graham}, {Heathcote}, {Lemoult},
  {Li}, {Mauclaire}, {Potter}, {Ribeiro}, {Matthews}, {Cameron}, {Guenther},
  {Kuschnig}, {Rowe}, {Rucinski}, {Sasselov}, \& {Weiss}}]{pablo2015}
{Pablo}, H., {Richardson}, N.~D., {Moffat}, A. F.~J., {et~al.} 2015, \apj, 809,
  134

\bibitem[{{Panchatsaram}(1981)}]{panchatsaram1981}
{Panchatsaram}, T. 1981, Bulletin of the Astronomical Society of India, 9, 139

\bibitem[{{Pavlovski} \& {Hensberge}(2005)}]{pavlovski2005}
{Pavlovski}, K. \& {Hensberge}, H. 2005, \aap, 439, 309

\bibitem[{{Pavlovski} \& {Hensberge}(2010)}]{pavlovski2010}
{Pavlovski}, K. \& {Hensberge}, H. 2010, in Astronomical Society of the Pacific
  Conference Series, Vol. 435, Binaries - Key to Comprehension of the Universe,
  ed. A.~{Pr{\v{s}}a} \& M.~{Zejda}, 207

\bibitem[{{Pavlovski} \& {Southworth}(2009)}]{pavlovski2009a}
{Pavlovski}, K. \& {Southworth}, J. 2009, \mnras, 394, 1519

\bibitem[{{Pavlovski} {et~al.}(2018){Pavlovski}, {Southworth}, \&
  {Tamajo}}]{pavlovski2018}
{Pavlovski}, K., {Southworth}, J., \& {Tamajo}, E. 2018, \mnras, 481, 3129

\bibitem[{{Pavlovski} {et~al.}(2009){Pavlovski}, {Tamajo}, {Koubsk{\'y}},
  {Southworth}, {Yang}, \& {Kolbas}}]{pavlovski2009b}
{Pavlovski}, K., {Tamajo}, E., {Koubsk{\'y}}, P., {et~al.} 2009, \mnras, 400,
  791

\bibitem[{{Paxton} {et~al.}(2011){Paxton}, {Bildsten}, {Dotter}, {Herwig},
  {Lesaffre}, \& {Timmes}}]{mesaI}
{Paxton}, B., {Bildsten}, L., {Dotter}, A., {et~al.} 2011, \apjs, 192, 3

\bibitem[{{Paxton} {et~al.}(2018){Paxton}, {Schwab}, {Bauer}, {Bildsten},
  {Blinnikov}, {Duffell}, {Farmer}, {Goldberg}, {Marchant}, {Sorokina},
  {Thoul}, {Townsend}, \& {Timmes}}]{mesaIV}
{Paxton}, B., {Schwab}, J., {Bauer}, E.~B., {et~al.} 2018, \apjs, 234, 34

\bibitem[{{Pietrzy{\'n}ski} {et~al.}(2013){Pietrzy{\'n}ski}, {Graczyk},
  {Gieren}, {Thompson}, {Pilecki}, {Udalski}, {Soszy{\'n}ski}, {Koz{\l}owski},
  {Konorski}, {Suchomska}, {Bono}, {Moroni}, {Villanova}, {Nardetto},
  {Bresolin}, {Kudritzki}, {Storm}, {Gallenne}, {Smolec}, {Minniti}, {Kubiak},
  {Szyma{\'n}ski}, {Poleski}, {Wyrzykowski}, {Ulaczyk}, {Pietrukowicz},
  {G{\'o}rski}, \& {Karczmarek}}]{pietrzynski2013}
{Pietrzy{\'n}ski}, G., {Graczyk}, D., {Gieren}, W., {et~al.} 2013, \nat, 495,
  76

\bibitem[{{Pols} {et~al.}(1997){Pols}, {Tout}, {Schroder}, \&
  {Eggleton}}]{pols1997}
{Pols}, O.~R., {Tout}, C.~A., {Schroder}, K.-P., \& {Eggleton}. 1997, \mnras,
  289, 869

\bibitem[{{Popper}(1974)}]{popper1974}
{Popper}, D.~M. 1974, \apj, 188, 559

\bibitem[{{Popper}(1980)}]{popper1980}
{Popper}, D.~M. 1980, ARA\&A, 18, 115

\bibitem[{{Popper} \& {Hill}(1991)}]{popper1991}
{Popper}, D.~M. \& {Hill}, G. 1991, \aj, 101, 600

\bibitem[{{Prsa} {et~al.}(2011){Prsa}, {Matijevic}, {Latkovic}, {Vilardell}, \&
  {Wils}}]{prsa2011}
{Prsa}, A., {Matijevic}, G., {Latkovic}, O., {Vilardell}, F., \& {Wils}, P.
  2011, {PHOEBE: PHysics Of Eclipsing BinariEs}, Astrophysics Source Code
  Library

\bibitem[{{Pr{\v s}a} \& {Zwitter}(2005)}]{prsa2005}
{Pr{\v s}a}, A. \& {Zwitter}, T. 2005, \apj, 628, 426

\bibitem[{{Raskin} {et~al.}(2011){Raskin}, {van Winckel}, {Hensberge},
  {Jorissen}, {Lehmann}, {Waelkens}, {Avila}, {de Cuyper}, {Degroote},
  {Dubosson}, {Dumortier}, {Fr{\'e}mat}, {Laux}, {Michaud}, {Morren}, {Perez
  Padilla}, {Pessemier}, {Prins}, {Smolders}, {van Eck}, \&
  {Winkler}}]{raskin2011}
{Raskin}, G., {van Winckel}, H., {Hensberge}, H., {et~al.} 2011, \aap, 526, A69

\bibitem[{{Reed}(1998)}]{reed1998}
{Reed}, B.~C. 1998, \jrasc, 92, 36

\bibitem[{{Ribas} {et~al.}(2000{\natexlab{a}}){Ribas}, {Guinan}, {Fitzpatrick},
  {DeWarf}, {Maloney}, {Maurone}, {Bradstreet}, {Gim{\'e}nez}, \&
  {Pritchard}}]{ribas2000c}
{Ribas}, I., {Guinan}, E.~F., {Fitzpatrick}, E.~L., {et~al.}
  2000{\natexlab{a}}, \apj, 528, 692

\bibitem[{{Ribas} {et~al.}(2000{\natexlab{b}}){Ribas}, {Jordi}, \&
  {Gim{\'e}nez}}]{ribas2000b}
{Ribas}, I., {Jordi}, C., \& {Gim{\'e}nez}, {\'A}. 2000{\natexlab{b}}, \mnras,
  318, L55

\bibitem[{{Ribas} {et~al.}(2000{\natexlab{c}}){Ribas}, {Jordi}, {Torra}, \&
  {Gim{\'e}nez}}]{ribas2000a}
{Ribas}, I., {Jordi}, C., {Torra}, J., \& {Gim{\'e}nez}, {\'A}.
  2000{\natexlab{c}}, \mnras, 313, 99

\bibitem[{{Ribas} {et~al.}(2005){Ribas}, {Jordi}, {Vilardell}, {Fitzpatrick},
  {Hilditch}, \& {Guinan}}]{ribas2005}
{Ribas}, I., {Jordi}, C., {Vilardell}, F., {et~al.} 2005, \apj, 635, L37

\bibitem[{{Ricker} {et~al.}(2015){Ricker}, {Winn}, {Vanderspek}, {Latham},
  {Bakos}, {Bean}, {Berta-Thompson}, {Brown}, {Buchhave}, {Butler}, {Butler},
  {Chaplin}, {Charbonneau}, {Christensen-Dalsgaard}, {Clampin}, {Deming},
  {Doty}, {De Lee}, {Dressing}, {Dunham}, {Endl}, {Fressin}, {Ge}, {Henning},
  {Holman}, {Howard}, {Ida}, {Jenkins}, {Jernigan}, {Johnson}, {Kaltenegger},
  {Kawai}, {Kjeldsen}, {Laughlin}, {Levine}, {Lin}, {Lissauer}, {MacQueen},
  {Marcy}, {McCullough}, {Morton}, {Narita}, {Paegert}, {Palle}, {Pepe},
  {Pepper}, {Quirrenbach}, {Rinehart}, {Sasselov}, {Sato}, {Seager},
  {Sozzetti}, {Stassun}, {Sullivan}, {Szentgyorgyi}, {Torres}, {Udry}, \&
  {Villasenor}}]{ricker2015}
{Ricker}, G.~R., {Winn}, J.~N., {Vanderspek}, R., {et~al.} 2015, Journal of
  Astronomical Telescopes, Instruments, and Systems, 1, 014003

\bibitem[{{Roxburgh}(1978)}]{roxburgh1978}
{Roxburgh}, I.~W. 1978, \aap, 65, 281

\bibitem[{{Roxburgh}(1992)}]{roxburgh1992}
{Roxburgh}, I.~W. 1992, \aap, 266, 291

\bibitem[{{Schmid} \& {Aerts}(2016)}]{schmid2016}
{Schmid}, V.~S. \& {Aerts}, C. 2016, \aap, 592, A116

\bibitem[{{Schmid} {et~al.}(2015){Schmid}, {Tkachenko}, {Aerts}, {Degroote},
  {Bloemen}, {Murphy}, {Van Reeth}, {P{\'a}pics}, {Bedding}, {Keen},
  {Pr{\v{s}}a}, {Menu}, {Debosscher}, {Hrudkov{\'a}}, {De Smedt}, {Lombaert},
  \& {N{\'e}meth}}]{schmid2015}
{Schmid}, V.~S., {Tkachenko}, A., {Aerts}, C., {et~al.} 2015, \aap, 584, A35

\bibitem[{{Schneider} {et~al.}(2014){Schneider}, {Langer}, {de Koter}, {Brott},
  {Izzard}, \& {Lau}}]{schneider2014}
{Schneider}, F.~R.~N., {Langer}, N., {de Koter}, A., {et~al.} 2014, \aap, 570,
  A66

\bibitem[{{Schroder} {et~al.}(1997){Schroder}, {Pols}, \&
  {Eggleton}}]{schroder1997}
{Schroder}, K.-P., {Pols}, O.~R., \& {Eggleton}, P.~P. 1997, \mnras, 285, 696

\bibitem[{{Shulyak} {et~al.}(2004){Shulyak}, {Tsymbal}, {Ryabchikova},
  {St{\"u}tz}, \& {Weiss}}]{shulyak2004}
{Shulyak}, D., {Tsymbal}, V., {Ryabchikova}, T., {St{\"u}tz}, C., \& {Weiss},
  W.~W. 2004, \aap, 428, 993

\bibitem[{{Simon} \& {Sturm}(1994)}]{simon1994}
{Simon}, K.~P. \& {Sturm}, E. 1994, \aap, 281, 286

\bibitem[{{Sim{\'o}n-D{\'\i}az} {et~al.}(2017){Sim{\'o}n-D{\'\i}az}, {Godart},
  {Castro}, {Herrero}, {Aerts}, {Puls}, {Telting}, \&
  {Grassitelli}}]{simondiaz2017}
{Sim{\'o}n-D{\'\i}az}, S., {Godart}, M., {Castro}, N., {et~al.} 2017, \aap,
  597, A22

\bibitem[{{Simonson} \& {van Someren Greve}(1976)}]{simonson1976}
{Simonson}, S.~C., I. \& {van Someren Greve}, H.~W. 1976, \aap, 49, 343

\bibitem[{{Stancliffe} {et~al.}(2015){Stancliffe}, {Fossati}, {Passy}, \&
  {Schneider}}]{stancliffe2015}
{Stancliffe}, R.~J., {Fossati}, L., {Passy}, J.~C., \& {Schneider}, F.~R.~N.
  2015, \aap, 575, A117

\bibitem[{{Stickland} {et~al.}(1992){Stickland}, {Koch}, \&
  {Pfeiffer}}]{strickland1992}
{Stickland}, D.~J., {Koch}, R.~H., \& {Pfeiffer}, R.~J. 1992, The Observatory,
  112, 277

\bibitem[{{Suchomska} {et~al.}(2019){Suchomska}, {Graczyk}, {Pietrzy{\'n}ski},
  {Gieren}, {Ostrowski}, {Smolec}, {Tkachenko}, {G{\'o}rski}, {Karczmarek},
  {Wielg{\'o}rski}, {Zgirski}, {Thompson}, {Villanova}, {Pilecki}, {Taormina},
  {Ko{\l}aczkowski}, {Narloch}, \& {Soszy{\'n}ski}}]{suchomska2019}
{Suchomska}, K., {Graczyk}, D., {Pietrzy{\'n}ski}, G., {et~al.} 2019, \aap,
  621, A93

\bibitem[{{Takata} {et~al.}(2012){Takata}, {Okazaki}, {Nagataki}, {Naito},
  {Kawachi}, {Lee}, {Mori}, {Hayasaki}, {Yamaguchi}, \& {Owocki}}]{takata2012}
{Takata}, J., {Okazaki}, A.~T., {Nagataki}, S., {et~al.} 2012, \apj, 750, 70

\bibitem[{{Tamajo} {et~al.}(2011){Tamajo}, {Pavlovski}, \&
  {Southworth}}]{tamajo2011}
{Tamajo}, E., {Pavlovski}, K., \& {Southworth}, J. 2011, \aap, 526, A76

\bibitem[{{Terrell}(1991)}]{terrell1991}
{Terrell}, D. 1991, \mnras, 250, 209

\bibitem[{{Tkachenko}(2015)}]{tkachenko2015}
{Tkachenko}, A. 2015, \aap, 581, A129

\bibitem[{{Tkachenko} {et~al.}(2014){Tkachenko}, {Degroote}, {Aerts},
  {Pavlovski}, {Southworth}, {P{\'a}pics}, {Moravveji}, {Kolbas}, {Tsymbal},
  {Debosscher}, \& {Cl{\'e}mer}}]{tkachenko2014}
{Tkachenko}, A., {Degroote}, P., {Aerts}, C., {et~al.} 2014, \mnras, 438, 3093

\bibitem[{{Torres} {et~al.}(2010){Torres}, {Andersen}, \&
  {Gim{\'e}nez}}]{torres2010}
{Torres}, G., {Andersen}, J., \& {Gim{\'e}nez}, A. 2010, ARA\&A, 18, 67

\bibitem[{{Torres} {et~al.}(2013){Torres}, {Ru{\'\i}z-Rodr{\'\i}guez},
  {Badenas}, {Prato}, {Schaefer}, {Wasserman}, {Mathieu}, \&
  {Latham}}]{torres2013}
{Torres}, G., {Ru{\'\i}z-Rodr{\'\i}guez}, D., {Badenas}, M., {et~al.} 2013,
  \apj, 773, 40

\bibitem[{{Torres} {et~al.}(2014{\natexlab{a}}){Torres}, {Sandberg Lacy},
  {Pavlovski}, {Feiden}, {Sabby}, {Bruntt}, \& {Viggo Clausen}}]{torres2014b}
{Torres}, G., {Sandberg Lacy}, C.~H., {Pavlovski}, K., {et~al.}
  2014{\natexlab{a}}, \apj, 797, 31

\bibitem[{{Torres} {et~al.}(2014{\natexlab{b}}){Torres}, {Vaz}, {Sandberg
  Lacy}, \& {Claret}}]{torres2014a}
{Torres}, G., {Vaz}, L. P.~R., {Sandberg Lacy}, C.~H., \& {Claret}, A.
  2014{\natexlab{b}}, \aj, 147, 36

\bibitem[{{Tsymbal}(1996)}]{tsymbal1996}
{Tsymbal}, V. 1996, in Astronomical Society of the Pacific Conference Series,
  Vol. 108, M.A.S.S., Model Atmospheres and Spectrum Synthesis, ed. S.~J.
  {Adelman}, F.~{Kupka}, \& W.~W. {Weiss}, 198

\bibitem[{{Valle} {et~al.}(2016){Valle}, {Dell'Omodarme}, {Prada Moroni}, \&
  {Degl'Innocenti}}]{valle2016}
{Valle}, G., {Dell'Omodarme}, M., {Prada Moroni}, P.~G., \& {Degl'Innocenti},
  S. 2016, \aap, 587, A16

\bibitem[{{Valle} {et~al.}(2017){Valle}, {Dell'Omodarme}, {Prada Moroni}, \&
  {Degl'Innocenti}}]{valle2017}
{Valle}, G., {Dell'Omodarme}, M., {Prada Moroni}, P.~G., \& {Degl'Innocenti},
  S. 2017, \aap, 600, A41

\bibitem[{{Valle} {et~al.}(2018){Valle}, {Dell'Omodarme}, {Prada Moroni}, \&
  {Degl'Innocenti}}]{valle2018}
{Valle}, G., {Dell'Omodarme}, M., {Prada Moroni}, P.~G., \& {Degl'Innocenti},
  S. 2018, \aap, 615, A62

\bibitem[{{van Leeuwen}(2007)}]{vanLeeuwen2007}
{van Leeuwen}, F. 2007, \aap, 474, 653

\bibitem[{{Van Reeth} {et~al.}(2016){Van Reeth}, {Tkachenko}, \&
  {Aerts}}]{vanReeth2016}
{Van Reeth}, T., {Tkachenko}, A., \& {Aerts}, C. 2016, \aap, 593, A120

\bibitem[{{Vaz} {et~al.}(2007){Vaz}, {Andersen}, \& {Claret}}]{vaz2007}
{Vaz}, L.~P.~R., {Andersen}, J., \& {Claret}, A. 2007, \aap, 469, 285

\bibitem[{{von Zeipel}(1924)}]{vonZeipel1924}
{von Zeipel}, H. 1924, \mnras, 84, 665

\bibitem[{{Wade} {et~al.}(2019){Wade}, {Smoker}, {Evans}, {Howarth}, {Barba},
  {Cox}, {Morrell}, {Naz{\'e}}, {Cami}, {Farhang}, {Walborn}, {Arias}, \&
  {Gamen}}]{wade2019}
{Wade}, G.~A., {Smoker}, J.~V., {Evans}, C.~J., {et~al.} 2019, \mnras, 483,
  2581

\bibitem[{{Wolf} {et~al.}(2002){Wolf}, {Harmanec}, {Diethelm}, {Hornoch}, \&
  {Eenens}}]{wolf2002}
{Wolf}, M., {Harmanec}, P., {Diethelm}, R., {Hornoch}, K., \& {Eenens}, P.
  2002, \aap, 383, 533

\bibitem[{{Wolf} {et~al.}(2006){Wolf}, {Ku{\v{c}}{\'a}kov{\'a}}, {Kolasa},
  {{\v{S}}tastn{\'y}}, {Bozkurt}, {Harmanec}, {Zejda}, {Br{\'a}t}, \&
  {Hornoch}}]{wolf2006}
{Wolf}, M., {Ku{\v{c}}{\'a}kov{\'a}}, H., {Kolasa}, M., {et~al.} 2006, \aap,
  456, 1077

\bibitem[{{Zahn}(1975)}]{zahn1975}
{Zahn}, J.~P. 1975, \aap, 41, 329

\bibitem[{{Zahn}(1977)}]{zahn1977}
{Zahn}, J.~P. 1977, {Penetrative convection in stars}, ed. E.~A. {Spiegel} \&
  J.~P. {Zahn}, 225--234

\bibitem[{{Zahn}(1991)}]{zahn1991}
{Zahn}, J.~P. 1991, \aap, 252, 179

\end{thebibliography}

\appendix

\section{CW Cep Marginalised posterior distributions}

\begin{figure}
\centering
\figps{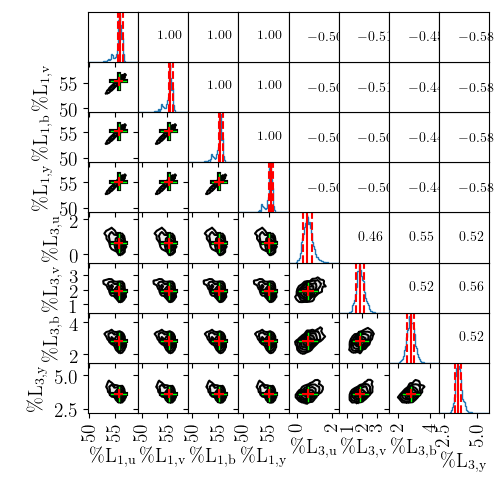}
\caption{Marginalised Posterior Distributions for primary and secondary passband luminosities for each observed filter. Median denoted by solid vertical red line, upper and lower bounds for 68.27\% CI denoted by dashed vertical red lines.}
\label{fig:cwcep_light_posteriors}
\end{figure}

\begin{figure}
\centering
\figps{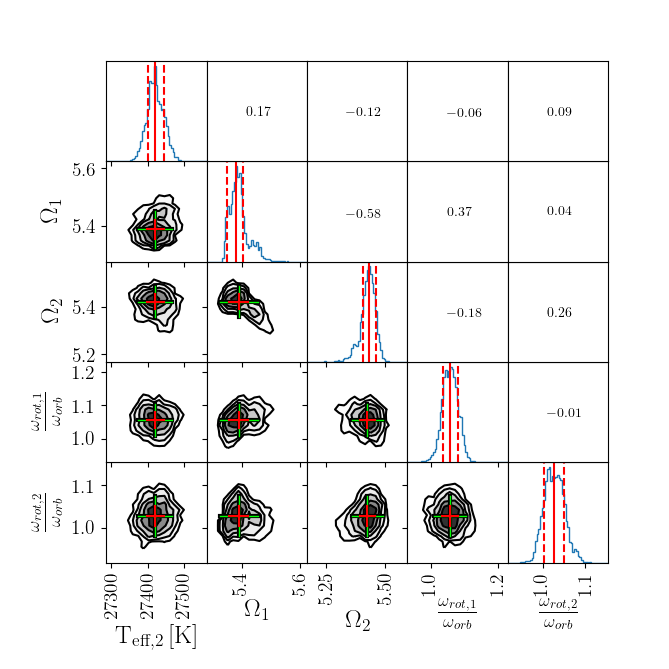}
\caption{Marginalised Posterior Distributions for primary and secondary parameters. Median denoted by solid vertical red line, upper and lower bounds for 68.27\% CI denoted by dashed vertical red lines.}
\label{fig:cwcep_star_posteriors}
\end{figure}

\begin{figure}
\centering
\figps{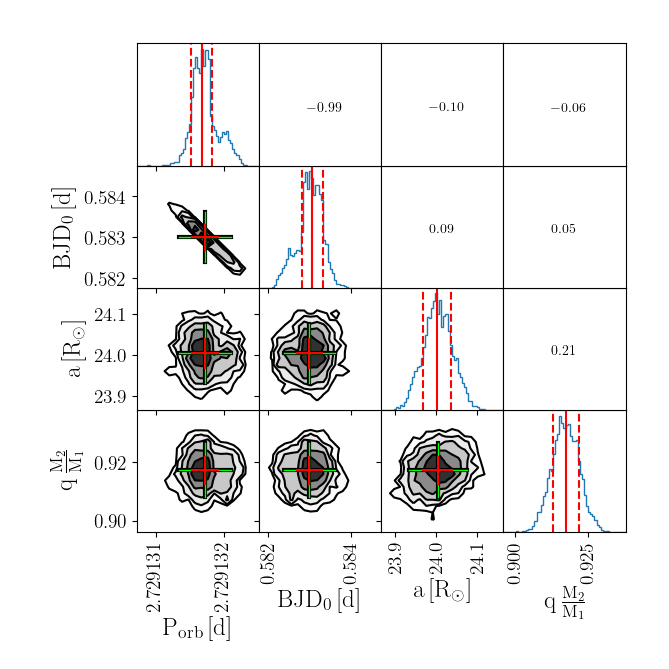}
\caption{Marginalised Posterior Distributions for orbital parameters. Median denoted by solid vertical red line, upper and lower bounds for 68.27\% CI denoted by dashed vertical red lines.}
\label{fig:cwcep_orbital_posteriors}
\end{figure}

\begin{figure}
\centering
\figps{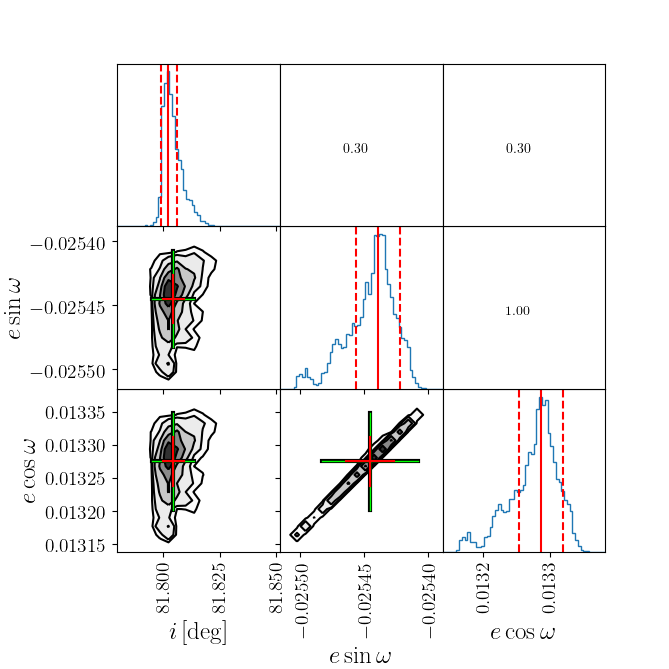}
\caption{Marginalised Posterior Distributions for system parameters. Median denoted by solid vertical red line, upper and lower bounds for 68.27\% CI denoted by dashed vertical red lines.}
\label{fig:cwcep_orientation_posteriors}
\end{figure}

\section{U Oph Marginalised posterior distributions}

\begin{figure}
\centering
\figps{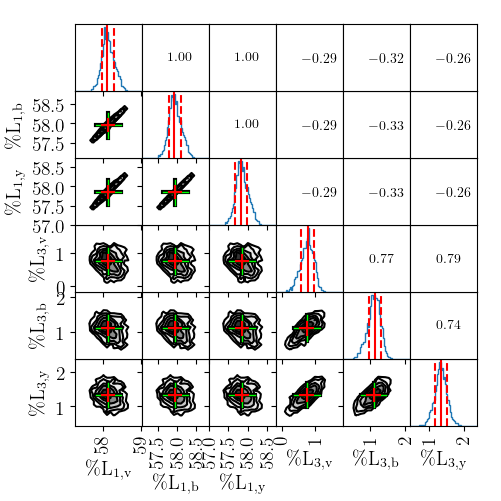}
\caption{Marginalised Posterior Distributions for primary and secondary passband luminosities for each observed filter. Median denoted by solid vertical red line, upper and lower bounds for 68.27\% CI denoted by dashed vertical red lines.}
\label{fig:uoph_light_posteriors}
\end{figure}

\begin{figure}
\centering
\figps{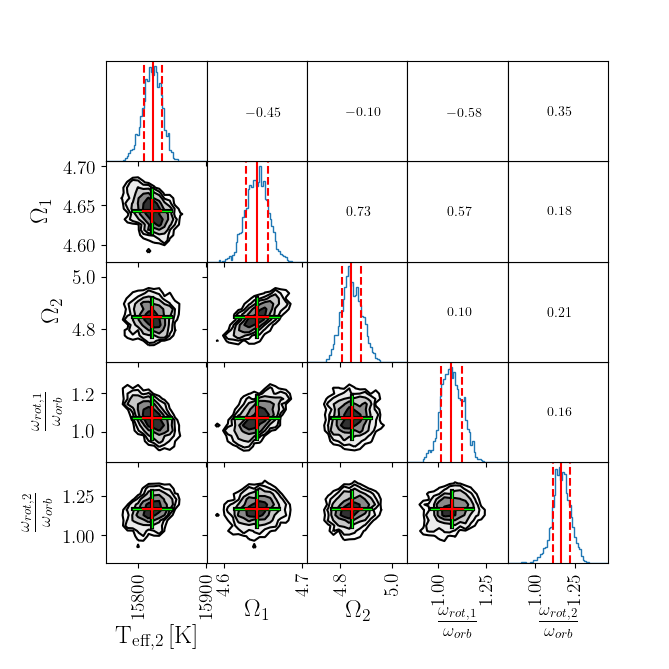}
\caption{Marginalised Posterior Distributions for primary and secondary parameters. Median denoted by solid vertical red line, upper and lower bounds for 68.27\% CI denoted by dashed vertical red lines.}
\label{fig:uoph_star_posteriors}
\end{figure}

\begin{figure}
\centering
\figps{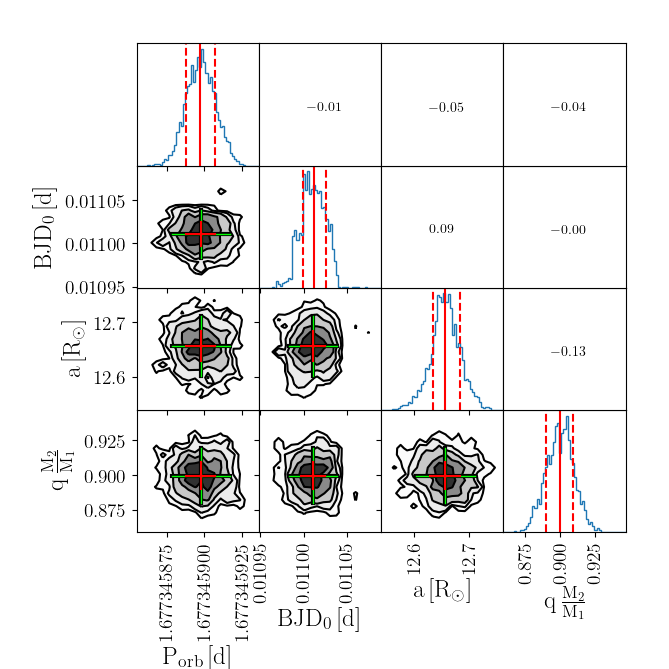}
\caption{Marginalised Posterior Distributions for orbital parameters. Median denoted by solid vertical red line, upper and lower bounds for 68.27\% CI denoted by dashed vertical red lines.}
\label{fig:uoph_orbital_posteriors}
\end{figure}

\begin{figure}
\centering
\figps{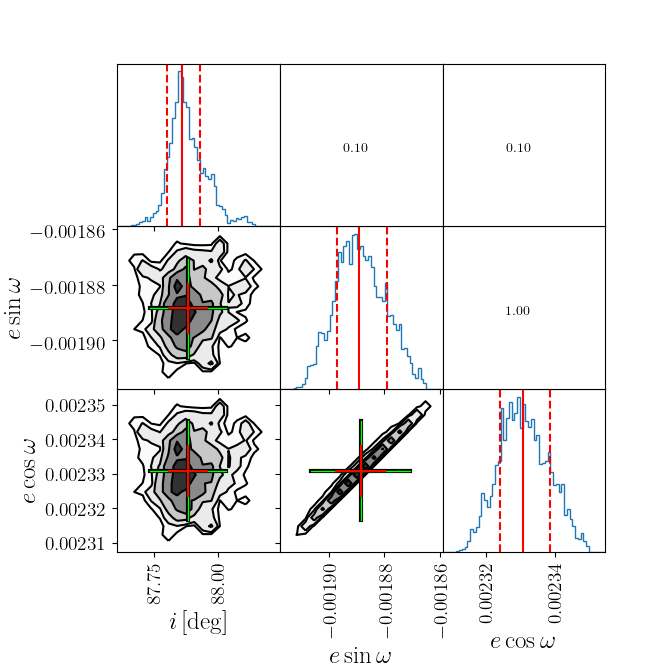}
\caption{Marginalised Posterior Distributions for system parameters. Median denoted by solid vertical red line, upper and lower bounds for 68.27\% CI denoted by dashed vertical red lines.}
\label{fig:uoph_orientation_posteriors}
\end{figure}

\label{lastpage}

\end{document}